\documentclass[twocolumn,superscriptaddress,showpacs,nofootinbib]{revtex4}
\usepackage{bm,color}

\usepackage{graphicx}
\usepackage{amsmath}
\usepackage{amssymb}
\usepackage{amsfonts}
\usepackage{bm}
\usepackage{color}

\def\be{\begin{equation}}
\def\ee{\end{equation}}
\def\bea{\begin{eqnarray}}
\def\eea{\end{eqnarray}}

\begin{document}

\title{Predictions from the logotropic model: the universal surface density of dark matter halos and the present proportion of dark matter and dark energy}

% First author block:
\author{Pierre-Henri Chavanis}
%\email{chavanis@irsamc.ups-tlse.fr}
% \homepage{An author's web page; optional}
\affiliation{Laboratoire de Physique
Th\'eorique, Universit\'e de Toulouse,
CNRS, UPS, France}
% You may list several affiliation, using separate commands for each:
%\affiliation{The third affiliation is shared by both co-authors}
%\author{Pierre-Henri Chavanis}
%\email{chavanis@irsamc.ups-tlse.fr}
%\affiliation{Laboratoire de Physique Th\'eorique, Universit\'e Paul
%Sabatier, 118 route de Narbonne 31062 Toulouse, France}
% For other authors please repeat the author block as needed
%\author{Second Author}
% Note how REVTeX 4 deals with identical affiliations
%\affiliation{The third affiliation is shared by both co-authors}

\begin{abstract}

The logotropic model  [P.H. Chavanis, Eur. Phys. J. Plus {\bf 130}, 130 (2015)]
may be an interesting 
alternative to the $\Lambda$CDM model. It is able to account for the present
accelerating expansion of the universe while solving at the same time the
core-cusp problem of the CDM model. In the logotropic model, there is a single
dark fluid. Its rest-mass plays the role of dark matter and its internal energy
plays the role of dark energy. We highlight two remarkable predictions of the
logotropic model. It yields cored dark matter halos with a universal surface
density equal to 
$\Sigma_0^{\rm th}=0.01955  c\sqrt{\Lambda}/G=133\, M_{\odot}/{\rm pc}^2$
without free parameter in very good 
agreement with the observational value $\Sigma_0^{\rm obs}=141_{-52}^{+83}\,
M_{\odot}/{\rm pc}^2$. It also predicts the present ratio of dark energy and
dark matter to be the pure number  $\Omega_{\rm de,0}^{\rm th}/\Omega_{\rm
dm,0}^{\rm th}=e=2.71828...$ in very good agreement with the observations giving
$\Omega_{\rm de,0}^{\rm obs}/\Omega_{\rm dm,0}^{\rm obs}=2.669\pm 0.08$.  Using 
the measured present proportion of baryonic matter $\Omega_{\rm b,0}^{\rm
obs}=0.0486\pm 0.0010$, we find that the values of the present proportion of
dark matter and dark energy are
$\Omega_{\rm dm,0}^{\rm th}=\frac{1}{1+e}(1-\Omega_{\rm b,0})=0.2559$ and
$\Omega_{\rm de,0}^{\rm th}=\frac{e}{1+e}(1-\Omega_{\rm b,0})=0.6955$ in very
good
agreement 
with the observational values
$\Omega_{\rm dm,0}^{\rm obs}=0.2589\pm 0.0057$ and
$\Omega_{\rm de,0}^{\rm obs}=0.6911\pm 0.0062$ within the error bars. These
theoretical 
predictions are obtained by advocating a mysterious strong cosmic coincidence
(dubbed ``dark
magic'') implying that our epoch plays a particular role in the history of the
universe.  We review the three types of logotropic models introduced in our
previous papers depending on whether the equation of state is expressed in terms
of the energy density, the rest-mass density, or the pseudo-rest mass density of
a complex scalar field. We discuss the similarities and the differences between
these models. Finally, we point out some intrinsic difficulties with the
logotropic model similar to those encoutered by the Chaplygin gas model and
discuss possible solutions.

\end{abstract}

% Insert suggested PACS numbers (up to 4) in braces.
% The PACS (Physics and Astronomy Classification Scheme)
% can be accessed on the web at http://www.aip.org/pacs/

% Insert keywords (up to about 4) in braces; optional.
% \keywords{Up to four keywords}

\maketitle

% Here the text of your article begins
%--------------------------------------------------------------------------------------------------------------------------------------------------

\section{Introduction}

Baryonic (visible) matter constitutes only $5\%$ of the content of the universe today. The rest
of the universe is made of approximately $25\%$ dark matter (DM) and  $70\%$
dark energy (DE) \cite{planck2014,planck2016}. DM can explain the flat rotation curves of the
spiral galaxies. It is also necessary to form the large-scale structures of the
universe. DE does not cluster but is responsible for the
late time acceleration of the universe revealed by the observations of type Ia
supernovae, the cosmic microwave background (CMB) anisotropies, and galaxy
clustering. Although there have been many theoretical attempts to explain DM and
DE, we still do not have a robust model for these dark components that can pass all the
theoretical and observational tests.

The most natural and simplest model
is the $\Lambda$CDM model which treats DM as a nonrelativistic cold pressureless
gas and DE as a cosmological constant $\Lambda$ (originally introduced by
Einstein \cite{einsteincosmo}) possibly representing vacuum energy
\cite{ss,carroll}. The effect of the cosmological constant is equivalent to
that
of a
fluid with a constant energy density $\epsilon_\Lambda=\Lambda c^2/8\pi G$ and a
negative pressure $P_\Lambda=-\epsilon_\Lambda$. Therefore, the $\Lambda$CDM
model 
is a two-fluid
model comprising  DM with an equation of state $P_{\rm dm}=0$ and DE with an equation of
state $P_{\rm de}=-\epsilon_{\rm de}$. When combined with the energy
conservation equation [see Eq. (\ref{econs})], the equation of state $P_{\rm
dm}=0$ implies that the DM density
decreases with the scale factor as $\epsilon_{\rm dm}=\epsilon_{\rm dm,0} a^{-3}$ and the
equation of state
$P_{\rm de}=-\epsilon_{\rm de}$ implies that the DE density is constant:
$\epsilon_{\rm de}=\epsilon_\Lambda$. Baryonic matter can also be modeled as a pressureless fluid ($P_{\rm b}=0$) whose density decreases as $\epsilon_{\rm b}=\epsilon_{\rm b,0} a^{-3}$. Therefore, the total energy density of the universe (baryons $+$ DM $+$ DE) evolves as
\begin{equation}
\epsilon=\frac{\epsilon_{\rm m,0}}{a^3}+\epsilon_\Lambda,
\label{intro1}
\end{equation}
where $\epsilon_{\rm m,0}=\epsilon_{\rm dm,0}+\epsilon_{\rm b,0}$ is the present density of (baryonic + dark) matter.  Matter dominates at early times when
the density is high ($\epsilon\sim \epsilon_{\rm m,0}/a^3$) and DE dominates at late times when the density is low ($\epsilon\rightarrow \epsilon_\Lambda$). The
scale factor increases algebraically as $a\propto t^{2/3}$ during the matter era (Einstein-de Sitter regime) and exponentially as $a\propto
{\rm exp}({\sqrt{\Lambda/3}\, t})$ during the DE era (de Sitter regime). As a
result, the universe undergoes a decelerated expansion followed 
by an accelerating expansion. At the present
epoch, both baryonic matter, DM and DE are important in the energy budget of the
universe. Introducing the Hubble constant $H=\dot a/a=(8\pi
G\epsilon/3c^2)^{1/2}$ 
[see Eq. (\ref{ak22})], we can rewrite Eq. (\ref{intro1}) as
\begin{equation}
\frac{H^2}{H_0^2}=\frac{\epsilon}{\epsilon_0}=\frac{\Omega_{\rm m,0}}{a^3}+\Omega_{\rm de,0},
\label{intro2}
\end{equation}
where $\epsilon_0=3H_0^2c^2/8\pi G$ is the present energy density of the
universe, $\Omega_{\rm m,0}=\epsilon_{\rm m,0}/\epsilon_0$ is the present
proportion of matter and $\Omega_{\rm de,0}=\epsilon_{\Lambda}/\epsilon_0$ is
the present proportion of DE. 
From the observations, we get $H_0=2.195\times 10^{-18}\, {\rm s}^{-1}$,
$\epsilon_0=7.75\times 10^{-7}\, {\rm g}\, {\rm m}^{-1}\,{\rm s}^{-2}$,
$\epsilon_0/c^2=8.62\times 10^{-24} {\rm g}\, {\rm m}^{-3}$,
$\Omega_{\rm b,0}=0.0486$, $\Omega_{\rm dm,0}=0.2589$  and  $\Omega_{\rm
de,0}=0.6911$. This gives $\epsilon_{\Lambda}=\Omega_{\rm
de,0}\epsilon_0=5.35\times 10^{-7}\, {\rm g}\, {\rm m}^{-1}\,{\rm s}^{-2}$.
Therefore, the value of the cosmological density
$\rho_{\Lambda}=\epsilon_{\Lambda}/c^2$ is
\begin{equation}
\rho_{\Lambda}=\frac{\Lambda}{8\pi G}=5.96\times 10^{-24} {\rm g}\, {\rm m}^{-3}
\label{intro3}
\end{equation}
and the value of the cosmological constant is $\Lambda=1.00\times 10^{-35}\, {\rm s}^{-2}$.

The CDM model faces important problems at the scale of DM halos such as the core-cusp problem
\cite{moore}, the missing satellite problem
\cite{satellites1,satellites2,satellites3}, 
and the ``too big to fail'' problem \cite{boylan}. This leads to the so-called
small-scale crisis of CDM 
\cite{crisis}. Basically, this is due to the assumption that DM is pressureless so there is 
nothing to balance the  gravitational attraction at high densities. As a
result, 
classical $N$-body simulations lead to DM halos exhibiting central cusps where
the density diverges as $r^{-1}$ \cite{nfw} while observations reveal that they
have constant density cores \cite{burkert}.  A possibility to solve these
problems is to take into account quantum mechanics. Fermionic and bosonic models
of DM halos display quantum cores, even at $T=0$, that replace the cusp (see,
e.g.,
\cite{modeldmB,modeldmF} and references therein). For self-gravitating
fermions, the
quantum pressure (leading to a fermion ball) is due to the Pauli exclusion
principle and for self-gravitating bosons the quantum pressure (leading to a
soliton) is  due to the Heisenberg uncertainty principle. However, these quantum
models cannot explain the observation that DM halos have a constant surface
density \cite{kormendy,spano,donato}
\begin{equation}
\Sigma_0=\rho_0 r_h=141_{-52}^{+83}\, M_{\odot}/{\rm pc}^2.
\label{intro4}
\end{equation}
Indeed, in fermionic and bosonic DM models,  the mass decreases  as 
the radius increases (see Appendix L of \cite{modeldmB})
instead of increasing according to $M_h\propto \Sigma_0 r_h^2$ as
implied by the constancy of the surface density.\footnote{In Refs.
\cite{modeldmB,modeldmF}, the law $M_h\propto \Sigma_0 r_h^2$ is heuristically
accounted for by the presence of an isothermal envelope (surrouding the quantum
core) whose temperature changes with $r_h$ according to $k_B T/m\propto
G\Sigma_0 r_h$.}

On the other hand, although the $\Lambda$CDM model is perfectly consistent with current
cosmological  observations, it faces two main problems. The first
problem is to explain the tiny value of the cosmological constant 
$\Lambda=1.00\times 10^{-35}\, {\rm s}^{-2}$.
Indeed, if DE can be attributed to vacuum fluctuations, quantum field
theory
predicts that $\Lambda$ should correspond to the Planck scale which is associated with the Planck density 
\begin{equation}
\rho_P=\frac{c^5}{\hbar G^2}=5.16\times 10^{99}\, {\rm g/m^3}.
\label{intro5}
\end{equation}
Now, the ratio between the Planck density (\ref{intro5}) and the cosmological density (\ref{intro3})  is
\begin{eqnarray}
\frac{\rho_P}{\rho_\Lambda}=\frac{8\pi c^5}{\hbar G \Lambda}\sim 10^{123}.
\label{intro6}
\end{eqnarray} 
Therefore, the observed cosmological constant  lies $123$ orders of magnitude
below the theoretical value. This is called the cosmological 
constant problem \cite{weinbergcosmo,paddycosmo}. The second problem is to
explain why DM and DE are of similar magnitudes today (within a factor $3$)
although they scale differently with the universe's expansion. This is the
cosmic coincidence problem \cite{stein1,zws},  which is a fine-tuning problem,
frequently triggering anthropic explanations.

For these reasons, other types of matter with negative pressure  that
can behave like a cosmological constant at late time have
been considered as candidates of DE: fluids of topological defects (domain
walls, cosmic strings...) \cite{vs,kt,fm,fg,pen}, $X$-fluids with a linear equation of
state $P=w_X\epsilon$ with a coefficient $w_X<-1/3$ triggering an accelerating
expansion of the universe \cite{tw,sw,csn}, a time-varying 
cosmological constant $\Lambda(t)$ \cite{bronstein,berto1,berto2,taha}, quintessence fields in the form of an evolving self-interacting scalar field (SF) minimally coupled to gravity
\cite{cds,pr,ratra,wett1,wett2,fhsw,cdf,fj,clw,fj2}, k-essence fields
corresponding to a SF with  
a noncanonical kinetic term \cite{chiba,ams,ams2} and even phantom or ghost
fields \cite{caldwell,ckw} represented by a SF with a negative kinetic term
implying that the energy density of the universe increases with the scale
factor. However, these models still face the cosmic coincidence problem because
they treat DM and DE as distinct entities.

Indeed, in the standard $\Lambda$CDM model and in the above-mentioned models, DM and DE 
are two independent components introduced to explain the clustering of matter and the
cosmic acceleration, respectively. However, DM and DE could be two different
manifestations of a single underlying substance (a dark fluid) called  
``quartessence'' \cite{makler}.  The most famous example is the Chaplygin gas
\cite{kmp} or generalized Chaplygin gas (GCG)  \cite{bentoGCG} in which the
pressure depends on a power of the density. The generalized Chaplygin equation
of state can be viewed as a polytropic equation of state with a negative
pressure \cite{cosmopoly1,cosmopoly2,cosmopoly3}
\begin{eqnarray}
P=K\rho^{\gamma}\qquad (K<0).
\label{intro7}
\end{eqnarray}
The original Chaplygin gas corresponds to $\gamma=-1$ \cite{kmp}. This dark
fluid behaves as DM at early times and as DE at late times. It provides
therefore a simple unification of DM and DE. This is an example of unified
dark matter and dark energy (UDM) model 
\cite{makler}. This dual behavior avoids fine-tuning problems since  the dark
fluid can be interpreted as an entangled mixture of DM and DE.  The $\Lambda$CDM
model can be seen as a UDM model where the pressure is a negative constant
\cite{avelinoZ,sandvik,cosmopoly2}
\begin{eqnarray}
P=-\rho_{\Lambda}c^2.
\label{intro8}
\end{eqnarray}
Indeed, by combining this equation of state with the energy conservation
equation we recover Eq. (\ref{intro1}) with just one fluid. This is a
particular case of the generalized Chaplygin gas corresponding to  $\gamma=0$
and $K=-\rho_{\Lambda}c^2$.

Recently, we have introduced the notion of logotropic dark fluid (LDF)
\cite{epjp,lettre,jcap,pdu,action,logosf} (see
also \cite{fa,cal1,ootsm,cal2,bal,mamon,bklmp,logogen})
where the pressure depends on the logarithm of the density as\footnote{The
logotropic equation of state can be obtained from the polytropic equation of
state (\ref{intro7}) in the limit $\gamma\rightarrow 0$ and $K\rightarrow
\infty$
with
$A=K\gamma$ finite \cite{logo,epjp,action,logosf}. It is interesting that the
Planck
density appears in this equation of state in order to make the argument of the
logarithm dimensionless.}
\begin{eqnarray}
P=A\ln \left (\frac{\rho}{\rho_P}\right ),
\label{intro9}
\end{eqnarray}
where  $\rho_P$ is the Planck density and $A$
is a new fundamental constant of physics superseding Einstein's cosmological
constant $\Lambda$. We will show that its value is given by
\begin{eqnarray}
A/c^2=\frac{\rho_{\Lambda}}{\ln\left(\frac{\rho_P}{\rho_{\Lambda}}\right )}=2.10\times 10^{-26}\, {\rm g}\, {\rm m}^{-3},
\label{intro10}
\end{eqnarray}
where $\rho_{\Lambda}$ is the cosmological density from Eq. (\ref{intro3}).
Therefore, the logotropic equation of state reads
\begin{eqnarray}
P=-\frac{\rho_{\Lambda}c^2}{\ln\left(\frac{\rho_P}{\rho_{\Lambda}}\right
)}\ln\left (\frac{\rho_P}{\rho}\right).
\label{intro11}
\end{eqnarray}
We note that $P=-\rho_{\Lambda}c^2$ when
$\rho=\rho_{\Lambda}$. It is convenient to write the fundamental constant $A$ under the form
\begin{eqnarray}
A=B\rho_\Lambda c^2,
\label{intro12}
\end{eqnarray}
where $B$ is the dimensionless number
\begin{eqnarray}
B=\frac{1}{\ln\left(\frac{\rho_P}{\rho_{\Lambda}}\right )}=3.53\times 10^{-3}.
\label{intro13}
\end{eqnarray}
Rewriting Eq. (\ref{intro13}) as  
\begin{eqnarray}
\frac{\rho_P}{\rho_\Lambda}=e^{1/B}
\label{intro14}
\end{eqnarray} 
and comparing this expression with Eq. (\ref{intro6}), we see that $B\simeq
1/[123\,
\ln(10)]$ is essentially the
inverse of the famous number $123$ (up to a conversion factor from neperian to
decimal logarithm). We note that $B$ has a small but {\it nonzero} value. This
is because $B$
depends on the Planck constant $\hbar$ through the Planck density  $\rho_P$
in Eq. (\ref{intro13}) and because $\hbar$ has a small but
nonzero value. In the classical (nonquantum) limit
$\hbar\rightarrow 0$, we find that $\rho_P\rightarrow +\infty$ and $B\rightarrow
0$. In that case, we recover the $\Lambda$CDM model. Indeed, when
$\rho_P\rightarrow +\infty$, 
the logotropic equation of state (\ref{intro11}) reduces to the constant
equation of state (\ref{intro8}) \cite{epjp,lettre,jcap,pdu,action,logosf}. The
fact that
$B$ is nonzero means that quantum effects ($\hbar\neq 0$) play
a fundamental role in the logotropic model. Since the effects of
$B$ manifest themselves in the late universe, this
implies (surprisingly!) that quantum mechanics affects the
late acceleration of the universe.

For a UDM model the equation of state can be specified in different manners
depending on whether the pressure $P$  is expressed in terms of the energy
density $\epsilon$ (model of type I), the rest-mass density $\rho_{\rm dm}=nm$
(model of type II), or the pseudo-rest mass density
$\rho=(m^2/\hbar^2)|\varphi|^2$  associated with a complex SF (model of type
III). In the nonrelativistic regime, these three formulations coincide and
$\rho=\rho_{\rm dm}=\epsilon/c^2$ represents the mass density. However, in the
relativistic regime, they lead to different models. The relation 
between these different models has been discussed in detail in \cite{action}.
For the logotropic equation of state, each of these models has been studied
exhaustively in a specific paper (the logotropic model of type II has been
discussed in \cite{epjp,lettre,jcap,pdu}  and the logotropic
models of type I and III have been discussed in
\cite{logosf}).  In the present paper, we provide
a brief comparison between these models and stress their main properties. We
also emphasize the main predictions of the logotropic
model \cite{epjp,lettre,jcap,pdu,action,logosf}:

(i) At small (galactic) scales, the logotropic model is able to solve the small-scale crisis of the CDM model. Indeed, contrary to the pressureless CDM model, the logotropic equation of state provides a pressure gradient that can balance the gravitational attraction and prevent gravitational collapse. As a result, logotropic DM halos present a central core rather than a cusp, in agreement with the observations. 

(ii) Remarkably, the logotropic model implies that DM halos have a
constant surface density  and it predicts its universal value $\Sigma_0^{\rm
th}=0.01955  c\sqrt{\Lambda}/G=133\, M_{\odot}/{\rm pc}^2$
\cite{epjp,lettre,jcap,pdu,action,logosf} without
adjustable parameter (here $\Lambda=1.00\times 10^{-35}\, {\rm
s}^{-2}$ is interpreted as an effective cosmological constant). This
theoretical value is in good agreement with the value $\Sigma_0^{\rm
obs}=141_{-52}^{+83}\, M_{\odot}/{\rm pc}^2$ obtained from the observations
\cite{donato}. 

(iii) As a corollary, the logotropic model implies that the mass of dwarf galaxies enclosed within a
sphere of fixed radius
$r_{u}=300\, {\rm pc}$ has a universal value
$M_{300}^{\rm th}=1.82\times 10^{7}\, M_{\odot}$, i.e. $\log
(M_{300}^{\rm th}/M_{\odot})=7.26$, in agreement with the
observations giving  $\log
(M_{300}^{\rm obs}/M_{\odot})=7.0^{+0.3}_{-0.4}$ \cite{strigari}.
The logotropic model also reproduces the Tully-Fisher
relation $M_{\rm b}\propto v_h^4$, where $M_{\rm b}$ is the baryonic mass and $v_h$ the
circular velocity at the halo radius, and predicts a value of the ratio
$(M_{\rm b}/v_h^4)^{\rm th}=46.4\,
M_{\odot}{\rm km}^{-4}{\rm s}^4$ which is close to the
observed value 
$(M_{\rm b}/v_h^4)^{\rm obs}=47\pm 6 \, M_{\odot}{\rm km}^{-4}{\rm s}^4$
\cite{mcgaugh}.

(iv) At large (cosmological) scales, the logotropic model is able to account for
the transition between DM and DE and for the present acceleration of the
universe.  Remarkably, it predicts the present ratio of DE and
DM to be the pure number  $\Omega_{\rm de,0}^{\rm th}/\Omega_{\rm dm,0}^{\rm
th}=e=2.71828...$ in very good agreement with the 
observations giving $\Omega_{\rm de,0}^{\rm obs}/\Omega_{\rm dm,0}^{\rm
obs}=2.669\pm 0.08$.  This then yields the values of
the present proportion of DM and DE as
$\Omega_{\rm dm,0}^{\rm th}=\frac{1}{1+e}(1-\Omega_{\rm b,0})=0.2559$ and
$\Omega_{\rm de,0}^{\rm th}=\frac{e}{1+e}(1-\Omega_{\rm b,0})=0.6955$  (where 
we have used $\Omega_{\rm b,0}^{\rm obs}=0.0486\pm 0.0010$)
\cite{pdu,logosf}, in
very good agreement with the observed values
$\Omega_{\rm dm,0}^{\rm obs}=0.2589\pm 0.0057$ and
$\Omega_{\rm de,0}^{\rm obs}=0.6911\pm 0.0062$ within the error bars.

\section{Logotropic equation of state of type I}
\label{sec_mtu}

In this section, we consider a relativistic  barotropic fluid described by an equation of
state of type I where the pressure $P=P(\epsilon)$ is specified as a function of
the energy density. 

\subsection{Friedmann equations}
\label{sec_mtuf}

We consider an expanding homogeneous universe and adopt the
Friedmann-Lema\^itre-Robertson-Walker (FLRW) metric. In that case, the Einstein field
equations reduce to the Friedmann equations \cite{weinbergbook}
\begin{equation}
\label{ak22}
H^2=\frac{8\pi
G}{3c^2}\epsilon,
\end{equation}
\begin{equation}
\label{ak22b}
2\dot H+3H^2=-\frac{8\pi
G}{c^2} P,
\end{equation}
where $H=\dot a/a$ is  the Hubble constant and $a(t)$ is the scale factor. To obtain Eq. (\ref{ak22}), we have
assumed that the universe is  flat ($k=0$) in agreement with the inflation
paradigm \cite{guthinflation} and the observations of the cosmic microwave
background (CMB) \cite{planck2014,planck2016}. On the other hand, we have set
the true cosmological constant   to zero ($\Lambda_{\rm true}=0$) since, in our
model, DE will be taken
into account in the equation of state $P(\epsilon)$. Equation (\ref{ak22b}) can also be written as
\begin{equation}
\label{ak22c}
\frac{\ddot a}{a}=-\frac{4\pi G}{3c^2}(3P+\epsilon),
\end{equation}
showing that the expansion of the universe is decelerating when $P>-\epsilon/3$
and accelerating when $P<-\epsilon/3$.

\subsection{Energy conservation equation}
\label{sec_mtue}

Combining Eqs. (\ref{ak22}) and (\ref{ak22b}), we obtain the energy conservation
equation
\begin{equation}
\label{econs}
\frac{d\epsilon}{dt}+3H\left
(\epsilon+P\right )=0.
\end{equation}
The energy density
increases with the scale factor when $P>-\epsilon$ and decreases with the scale
factor when $P<-\epsilon$. The latter case corresponds to a phantom
behavior \cite{caldwell,ckw}.

For a given equation of state $P(\epsilon)$ we can solve Eq. (\ref{econs}) to get 
\begin{equation}
\label{econsint}
\ln a=-\frac{1}{3}\int \frac{d\epsilon}{\epsilon+P(\epsilon)}.
\end{equation}
This equation determines the relation between the energy  density $\epsilon$ and the scale factor $a$. We can then solve the Friedmann equation
(\ref{ak22}) with $\epsilon(a)$  to obtain the temporal evolution 
of the scale factor $a(t)$. We note that the function $a(\epsilon)$ is univalued. As a result, an 
equation of state of type I describes either a normal behavior or
a phantom behavior but it cannot
describe the transition from a normal to a phantom behavior.

\subsection{Logotropic equation of state}
\label{sec_mtul}

For the logotropic equation of state of type I:
\begin{eqnarray}
P=A\ln\left (\frac{\epsilon}{\rho_P c^2}\right ),
\label{lu1}
\end{eqnarray}
the energy conservation equation (\ref{econsint}) reads
\begin{equation}
\label{dim1}
\ln a=-\frac{1}{3}\int_{\epsilon_0}^{\epsilon}
\frac{d\epsilon'}{\epsilon'+A\ln\left (\frac{\epsilon'}{\rho_P c^2}\right )},
\end{equation}
where $\epsilon_0$ denotes the present energy density of the universe (when
$a=1$).  This equation determines the evolution of the energy density
$\epsilon(a)$ as a function of the scale factor. When $a\rightarrow 0$, Eq. (\ref{dim1}) reduces to
\begin{equation}
\label{dim1b}
\ln a\sim -\frac{1}{3}\int^{\epsilon}
\frac{d\epsilon'}{\epsilon'},
\end{equation}
implying that the energy density decreases like
$\epsilon\propto a^{-3}$ as the universe expands. This corresponds to the  DM regime. The scale factor increases in time like $a\propto t^{2/3}$ (Einstein-de Sitter). When $a\rightarrow +\infty$, the energy density tends to a constant $\epsilon_{\rm min}$ which is the solution of the equation 
\begin{equation}
\epsilon_{\rm
min}+A\ln\left (\frac{\epsilon_{\rm min}}{\rho_P c^2}\right )=0.
\label{dim1c}
\end{equation}
This corresponds to the DE regime. The scale factor increases exponentially rapidly in time like $a\propto {\rm exp}(\sqrt{8\pi G\epsilon_{\rm min}/3c^2}\, t)$  (de Sitter).

\subsection{The value of $A$}

The behavior of the energy density in the logotropic model of type I is similar to the one in the $\Lambda$CDM model. If we identify $\epsilon_{\rm min}$ with the DE density $\rho_{\Lambda}c^2$ in the $\Lambda$CDM
model, which is equal to the asymptotic value of $\epsilon$ for $a\rightarrow
+\infty$, we get
\begin{eqnarray}
A=\frac{\rho_{\Lambda}c^2}{\ln\left(\frac{\rho_P}{\rho_{\Lambda}}\right )}.
\label{dim2}
\end{eqnarray}
This determines the value of the constant $A$ in the logotropic model as given by Eq. (\ref{intro10}). We then find that the pressure decreases monotonically from $+\infty$ to $P_{\rm min}=-\epsilon_{\Lambda}$ as the universe expands. The pressure $P$ is positive when $\epsilon>\rho_Pc^2$ and negative
when $\epsilon<\rho_Pc^2$. It vanishes at $\epsilon=\rho_Pc^2$. Since the logotropic model is a unification of DM and DE, it is not expected to be valid in the early universe. Therefore, the pressure is always negative in the regime of interest ($\epsilon\ll \rho_Pc^2$) where the logotropic model is valid.

\subsection{Evolution of the universe}

\begin{figure}[!h]
\begin{center}
\includegraphics[clip,scale=0.3]{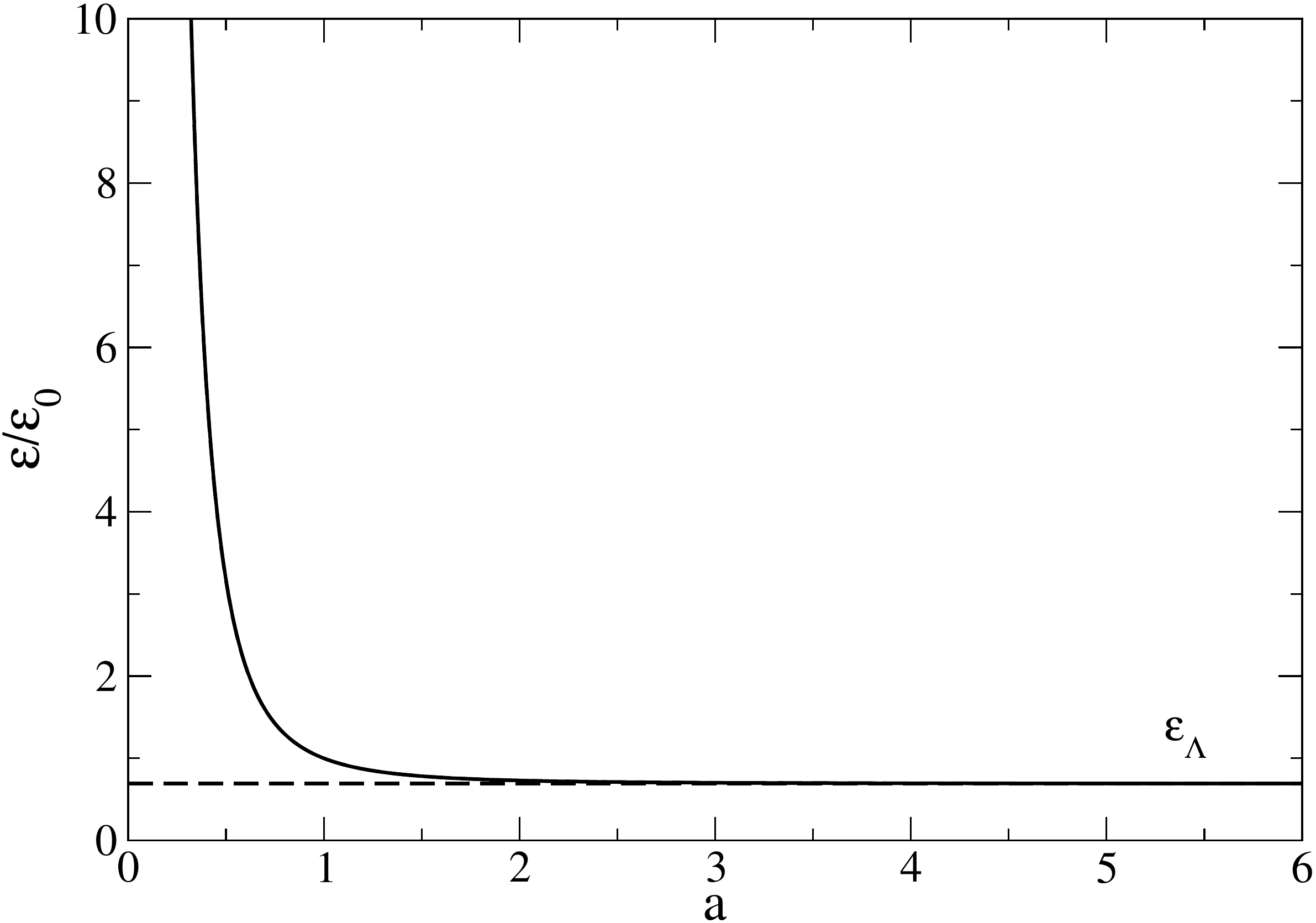}
\caption{Normalized energy density $\epsilon/\epsilon_0$ as a function of the
scale factor $a$ for the logotropic model of type I. It is compared with the
$\Lambda$CDM model. The two curves are indistinguishable on the figure.}
\label{aepsN}
\end{center}
\end{figure}

Setting $x=\epsilon'/\epsilon_0$ and $A=B \, \Omega_{\rm de,0} \epsilon_0$ we can rewrite Eq.
(\ref{dim1}) as
\begin{equation}
\label{dim3}
\ln a=-\frac{1}{3}\int_{1}^{\epsilon/\epsilon_0}
\frac{dx}{x+B\Omega_{\rm de,0}\left (\ln
x-\ln\Omega_{\rm de,0}-\frac{1}{B}\right )}.
\end{equation}
The function $\epsilon/\epsilon_0(a)$ is plotted in Fig. \ref{aepsN}. We have
taken $\Omega_{\rm de,0}=0.6911$ and $B=3.53\times
10^{-3}$. The logotropic
model of type I behaves similarly to the $\Lambda$CDM
model.  The two models coincide in the limit $B\rightarrow 0$ where  Eq.
(\ref{dim3}) returns Eq. (\ref{intro2}). For the simplicity of the presentation,
we have ignored here the contribution of baryonic matter but it is 
straightforward to take it into account.

\section{Logotropic equation of state of type II}
\label{sec:cosm}

In this section, we consider a relativistic barotropic fluid  described by an equation of state of type II where the pressure $P=P(\rho_{\rm dm})$ is specified as a function of the rest-mass density.  The notation $\rho_{\rm dm}$ for the rest-mass density will soon become clear. 

\subsection{First principle of thermodynamics}

The first principle of
thermodynamics for a relativistic gas can be
written as
\begin{equation}
\label{mtd1}
d\left (\frac{\epsilon}{\rho_{\rm dm}}\right )=-Pd\left
(\frac{1}{\rho_{\rm dm}}\right )+Td\left (\frac{s}{\rho_{\rm dm}}\right ),
\end{equation}
where
\begin{eqnarray}
\label{mtd2}
\epsilon=\rho_{\rm dm} c^2+u
\end{eqnarray}
is the energy density including the rest-mass energy density $\rho_{\rm dm} c^2$ (where
$\rho_{\rm dm}=n m$ 
is the rest-mass density) and the internal energy density $u$, $s$ is
the entropy density, $P$ is the pressure, and $T$ is the temperature. We assume that
$Td(s/\rho_{\rm dm})=0$. This corresponds
to a cold ($T=0$) or isentropic ($s/\rho_{\rm dm}={\rm cst}$) gas. In that
case, Eq.
(\ref{mtd1}) reduces to
\begin{equation}
\label{mtd3}
d\left (\frac{\epsilon}{\rho_{\rm dm}}\right )=-Pd\left
(\frac{1}{\rho_{\rm dm}}\right )=\frac{P}{\rho_{\rm dm}^2}\, d\rho_{\rm dm}.
\end{equation}
Assuming that $P=P(\rho_{\rm dm})$ and integrating Eq. (\ref{mtd3}), we obtain Eq. (\ref{mtd2}) with
\begin{eqnarray}
\label{mtd6}
u(\rho_{\rm dm})=\rho_{\rm dm} \int^{\rho_{\rm dm}} \frac{P
\left(\rho'\right)}{\rho{'}^2} d \rho'.
\end{eqnarray}
This relation determines the internal energy as a function of the equation of
state $P(\rho_{\rm dm})$. Inversely, the equation of state is determined by the
internal energy $u(\rho_{\rm dm})$ through the relation
\begin{equation}
\label{mtd7}
P(\rho_{\rm dm})=\rho_{\rm dm} u'(\rho_{\rm dm})-u(\rho_{\rm dm}).
\end{equation}

Let us apply these equations in a cosmological context, namely for a spatially  homogeneous
fluid in an expanding background. Combining the energy conservation equation (\ref{econs}) with the first principle of thermodynamics (\ref{mtd3}) which can be rewritten as
\begin{eqnarray}
d \epsilon =  \frac{P + \epsilon}{\rho_{\rm dm}} d \rho_{\rm dm},
\label{eq10}
\end{eqnarray}
we find that the rest-mass density satisfies the
equation \cite{epjp,lettre}
\begin{eqnarray}
\frac{d\rho_{\rm dm}}{dt}+3H\rho_{\rm dm}=0.
\label{eq15a}
\end{eqnarray}
This equation can be integrated into
\begin{eqnarray}
\rho_{\rm dm}  = 
\frac{\rho_{{\rm dm},0}}{a^3},
\label{eq15b}
\end{eqnarray}
where $\rho_{{\rm dm},0}$ is the present value of the
rest-mass density. Equations (\ref{eq15a}) and (\ref{eq15b}) express the
conservation 
of the rest-mass of the dark fluid. As argued in our previous
papers \cite{epjp,lettre},  the
rest-mass energy density $\rho_{\rm dm} c^2$
plays the role of DM and the internal
energy $u$ plays the role of DE. Therefore, we have 
\begin{eqnarray}
\epsilon_{\rm dm}  =  \rho_{\rm dm} c^2= \frac{\Omega_{{\rm dm},0}\epsilon_0}{a^{3}}
\label{eq16}
\end{eqnarray}
and
 \begin{eqnarray}
\label{mtd2b}
\epsilon_{\rm de}=u(\rho_{\rm dm})=u\left (\frac{\Omega_{\rm
dm,0}\epsilon_0}{c^2a^3} \right ).
\end{eqnarray}
The total energy density of the dark fluid then reads
\begin{eqnarray}
\label{mtd2bb}
\epsilon=\frac{\Omega_{\rm dm,0}\epsilon_0}{a^3}+u\left (\frac{\Omega_{\rm
dm,0}\epsilon_0}{c^2a^3} \right ).
\end{eqnarray}
The decomposition $\epsilon=\rho_{\rm dm} c^2+u=\epsilon_{\rm dm}+\epsilon_{\rm
de}$ provides a simple and nice interpretation of DM and DE in terms of the
rest-mass energy and 
internal energy of a single DF \cite{epjp,lettre}. For given $P(\rho_{\rm
dm})$
or
$u(\rho_{\rm dm})$ the relation between the energy density and the
scale factor is determined by Eq.  (\ref{mtd2bb}). We can then solve
the Friedmann equation (\ref{ak22}) with $\epsilon(a)$ to determine the temporal evolution of the
scale factor $a(t)$.

{\it Remark:} The equation of state parameter $w=P/\epsilon$ is given by 
\begin{eqnarray}
w=\frac{\rho_{\rm dm} u'(\rho_{\rm dm})-u(\rho_{\rm dm})}{\rho_{\rm dm}
c^2+u(\rho_{\rm dm})}.
\label{rtf7w}
\end{eqnarray}
For a barotropic equation 
of state of type II the universe exhibits a normal behavior ($w>-1$) when
$1+(1/c^2)u'(\rho_{\rm dm})>0$ and a phantom behavior ($w<-1$)  when
$1+(1/c^2)u'(\rho_{\rm dm})<0$. An equation of state of type II can describe
the transition from a normal to a phantom behavior.

\subsection{Logotropic equation of state and logarithmic internal energy}

For the logotropic equation of state of type II
\cite{epjp,lettre}
\begin{eqnarray}
P  =    A \ln \left(\frac{\rho_{\rm dm}}{\rho_{P}} \right),
\label{eq1bo}
\end{eqnarray}
the internal energy obtained from Eq. (\ref{mtd6}) reads
\begin{eqnarray}
u=-A\left\lbrack 1+\ln\left (\frac{\rho_{\rm dm}}{\rho_P}\right
)\right\rbrack.
\label{nrj}
\end{eqnarray}
Therefore, the energy density of the LDF is 
\begin{eqnarray}
\epsilon=\rho_{\rm dm} c^2-A\left\lbrack 1+\ln\left (\frac{\rho_{\rm dm}}{\rho_P}\right
)\right\rbrack=\epsilon_{\rm dm}+\epsilon_{\rm de},
\label{nrja}
\end{eqnarray}
where the first term (rest-mass)  is interpreted as DM and the second term
(internal energy) as DE.  Our model
provides a simple unification of these two entities. Eliminating $\rho_{\rm
dm}$ between Eqs. (\ref{eq1bo})
and (\ref{nrja}), the equation of state $P(\epsilon)$ is
given in the reversed form $\epsilon(P)$ by
\begin{eqnarray}
\epsilon=e^{P/A} \rho_P c^2-P-A.
\end{eqnarray}
This is the equation of state of type I corresponding to the logotropic model of
type II \cite{action}.
Following \cite{epjp,lettre}, we define the dimensionless
parameter $B$ through the relation
\begin{eqnarray}
\frac{\rho_P}{\rho_{{\rm dm},0}}  =  e^{1+ 1/B}.
\label{eq18}
\end{eqnarray}
Using Eqs. (\ref{eq15b}) and (\ref{eq18}), the pressure and the total energy
density can be expressed as a function of the scale factor as
\begin{eqnarray}
P  =  -  A \left (1+\frac{1}{B}+3\ln a\right ),
\label{eq1co}
\end{eqnarray}
\begin{eqnarray}
\epsilon  = \frac{\Omega_{{\rm dm},0} \epsilon_0}{a^3} + A \left (\frac{1}{B}+3\ln
a\right ).
\label{eq22o}
\end{eqnarray}

\subsection{The value of $A$}
\label{sec_b}

According to Eqs. (\ref{nrja}) and (\ref{eq22o}) the DE density is given by
\begin{eqnarray}
\epsilon_{\rm de}=A \left (\frac{1}{B}+3\ln a\right ).
\label{nrjb}
\end{eqnarray}
Applying this relation at the present time ($a=1$), we obtain the relation
\begin{eqnarray}
A  =  B \Omega_{{\rm de},0}  \epsilon_0.
\label{eq23g}
\end{eqnarray}
Explicating the expression of $B$ from Eq. (\ref{eq18}) we get
\begin{eqnarray}
A=\frac{\Omega_{{\rm de},0}\epsilon_0}{
\ln\left
(\frac{\rho_P c^2}{\Omega_{{\rm dm},0}\epsilon_0}\right
)-1}.
\label{f7}
\end{eqnarray}
If we view $A$ as a fundamental constant of physics, 
this equation gives a relation between the present fraction $\Omega_{{\rm
dm},0}$ of DM, the present fraction $\Omega_{{\rm de},0}$ of DE and the present
energy density $\epsilon_0$. Inversely, we can use the measured values of
$\Omega_{{\rm dm},0}$, $\Omega_{{\rm de},0}$ and $\epsilon_0$ to determine the
expression of the constant $A$ appearing in the logotropic equation of state.
Therefore, there is no free (undetermined) parameter in our model.

As in our previous papers \cite{epjp,lettre,jcap,pdu,action,logosf}, it is
convenient
to give a special name to the present density of DE and write it as
\begin{eqnarray}
\epsilon_\Lambda=\rho_{\Lambda}c^2=\Omega_{\rm
de,0}\epsilon_0.
\label{impid6}
\end{eqnarray}
In the $\Lambda$CDM model, the DE density $\rho_{\Lambda}=\Lambda/8\pi G$ is
constant. It represents the cosmological density which is determined by
Einstein's cosmological constant $\Lambda$. The present DE density of the
universe coincides with the
constant DE density of the universe in the  $\Lambda$CDM model. This justifies
the
notation from Eq. (\ref{impid6}). With this notation we can rewrite Eq.
(\ref{f7}) as
\begin{eqnarray}
A=\frac{\rho_{\Lambda}c^2}{
\ln\left
(\frac{\rho_P}{\rho_{\Lambda}}\right )+\ln\left (
\frac{\Omega_{{\rm de},0}}{\Omega_{{\rm dm},0}}\right
)-1}.
\label{f8}
\end{eqnarray}
Recalling the value of the ratio $\rho_P/\rho_\Lambda$ from Eq. (\ref{intro6}), 
we find that $\ln\left ({\rho_P}/{\rho_{\Lambda}}\right )=283$. On the other
hand, using the measured values of $\Omega_{{\rm de},0}^{\rm obs}=0.6911$ and
$\Omega_{{\rm dm},0}^{\rm obs}=0.2589$, we get $\ln\left (
{\Omega_{{\rm de},0}}/{\Omega_{{\rm dm},0}}\right
)-1=-0.0182$. Therefore, the first term in the denominator of Eq. (\ref{f8}) is
much larger than the second term so that, in very good approximation, we have
\begin{eqnarray}
A=\frac{\rho_{\Lambda}c^2}{\ln\left(\frac{\rho_P}{\rho_{\Lambda}}\right )},
\label{a2b}
\end{eqnarray}
as given by Eq. (\ref{intro10}). Similarly, $B$ is given in very good
approximation by Eq. (\ref{intro13}).

\subsection{Evolution of the universe}
\label{sec_arg}

Using Eq. (\ref{eq23g}), the pressure $P$ and the energy density $\epsilon$ can
be rewritten as
\begin{eqnarray}
P =  -  \Omega_{{\rm de},0}  \epsilon_0 \left (B+1+3B\ln a\right ),
\label{eq1cob}
\end{eqnarray}
\begin{eqnarray}
\frac{H^2}{H_0^2}=\frac{\epsilon}{\epsilon_0} =  \frac{\Omega_{{\rm dm},0}}{a^3} +
\Omega_{{\rm de},0}\left(
1+ 3 B \ln a \right).
\label{eq27}
\end{eqnarray}
In Eq. (\ref{eq27}) we have combined the relation $\epsilon(a)$ with the
Friedmann equation
(\ref{ak22}) to
obtain a  differential equation determining the temporal evolution of the scale
factor. If we take into account the presence of baryons, we must add a term
$\epsilon_{\rm b}=\Omega_{\rm b,0}\epsilon_0/{a^3}$ in the energy density.  The 
 $\Lambda$CDM model is recovered for $B=0$ corresponding to the limit
$\hbar\rightarrow 0$ or $\rho_P\rightarrow +\infty$.  In that case, the internal
energy is constant ($u=\epsilon_{\Lambda}$) and Eqs. (\ref{eq1cob}) and
(\ref{eq27}) return Eqs. (\ref{intro1}), (\ref{intro2}) and (\ref{intro8}). The
evolution of the universe in the logotropic model of type
II 
has been discussed in detail in \cite{epjp,lettre,jcap}.
Below, we just recall the main results of these studies.

\begin{figure}[!h]
\begin{center}
\includegraphics[clip,scale=0.3]{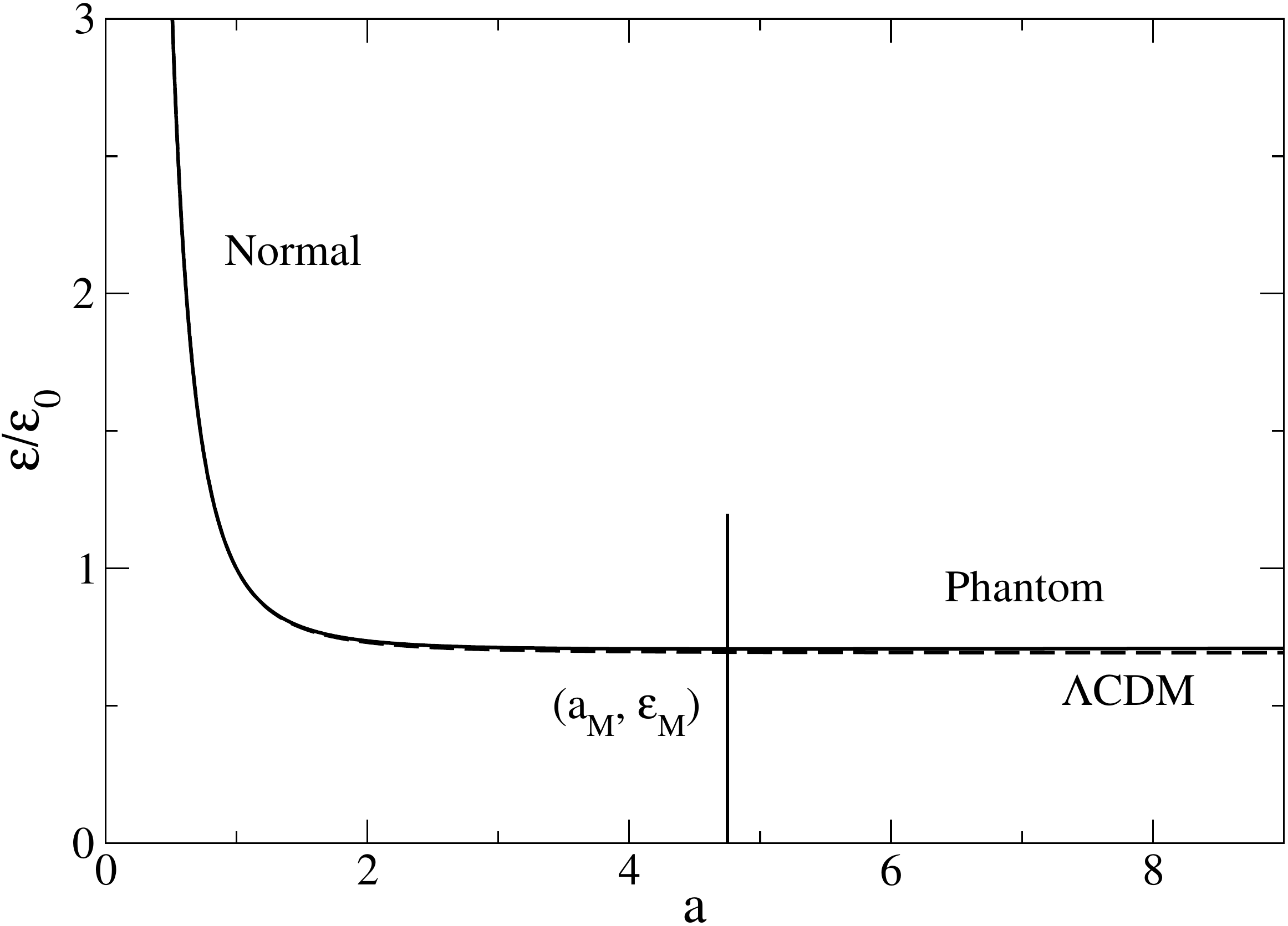}
\caption{Normalized energy density $\epsilon/\epsilon_0$ as a function of the
scale factor $a$ for the logotropic model of type II. It is compared with the
$\Lambda$CDM model. The two curves are indistinguishable up to about 
$a_M=5.01$ when the logotropic universe becomes phantom.}
\label{aepsT2}
\end{center}
\end{figure}

Starting from $+\infty$, the energy density $\epsilon$ first decreases, 
reaches a minimum $\epsilon_{M}=A\ln(\rho_P c^2/A)>0$ at $\rho_{\rm
dm}=A/c^2$, then increases to $+\infty$. The minimum energy density
$\epsilon_{M}=0.707\, \epsilon_0=1.02\, \epsilon_{\Lambda}$ is achieved at a
scale factor $a_M=5.01$
corresponding to a time $t_M=2.81\, H_0^{-1}=40.6 \, {\rm Gyrs}$ (the age of the
universe is $t_0=13.8\, {\rm Gyrs}$). At early times, the pressure is
negligible
with respect to the energy density and the LDF is equivalent
to a pressureless
fluid like in the CDM model. The energy density decreases as $\epsilon\sim
\Omega_{\rm dm,0}\epsilon_0/a^3$ and the scale factor increases algebraically as
$a\propto t^{2/3}$ (Einstein-de Sitter). This leads  to a decelerated expansion
of the universe.  At later times, the negative pressure of the LDF becomes
efficient and explains the acceleration of the universe that we observe today.
Ultimately, DE dominates over DM (more precisely the internal energy of the LDF
dominates over its rest-mass energy) and the energy density increases
logarithmically  as $\epsilon \sim \Omega_{\rm de,0}\epsilon_0(1+3B\ln a)$. The
scale factor has a super de Sitter behavior $a\propto {\rm exp}[3B\Omega_{\rm
de,0}H_0^2t^2/4]$ \cite{epjp,lettre,jcap}. This corresponds to a phantom regime
since the energy density increases as the universe expands. In the late
universe, we have $\epsilon \sim \Omega_{{\rm de},0} \epsilon_0 (1+3B\ln a)$ and
$P
\sim - \Omega_{{\rm de},0} \epsilon_0 (B+1+3B\ln a)$, which implies that the
equation
of state $P (\epsilon)$
behaves asymptotically as $P\sim -\epsilon-A$.

As the universe expands, the pressure decreases from $+\infty$ to $-\infty$. 
The pressure $P$ is positive when $\rho_{\rm dm}>\rho_P$ and negative
when $\rho_{\rm dm}<\rho_P$. It vanishes at $\rho_{\rm dm}=\rho_P$. Since the logotropic model is a unification of DM and DE, it is not expected to be valid in the early universe. Therefore, the pressure is always negative in the regime of interest ($\rho_{\rm dm}\ll \rho_P$) where the logotropic model is valid. 

The DE density $\epsilon_{\rm de}$ increases from
$-\infty$ to $+\infty$. The DE is negative when $\rho_{\rm dm}>\rho_P/e$ and positive
when  $\rho_{\rm dm}<\rho_P/e$  (its value at $\rho_{\rm dm}=\rho_P$ is
$\epsilon_{\rm de}=-A<0$). In the logotropic model, since the DE density corresponds to the internal energy density $u$ of the LDF, it can very well be negative as long as the total energy density
$\epsilon$ is positive. In the regime of interest
($\rho_{\rm dm}\ll\rho_P$) where the logotropic model is valid, the DE density $\epsilon_{\rm de}$ is positive.

The function $\epsilon/\epsilon_0(a)$ is plotted in
Fig. \ref{aepsT2}. We have 
taken $\Omega_{\rm m,0}=0.3075$, $\Omega_{\rm de,0}=0.6911$ and $B=3.53\times
10^{-3}$. The logotropic
model is able to account for the transition between a DM era where the expansion
of the universe is decelerated and a DE era where the expansion of the universe
is accelerating.  It is indistinguishable from the
$\Lambda$CDM model up to the present time for what concerns the evolution of the
cosmological background. The two models will differ in about $27$ Gyrs when the
logotropic model will start to exhibit a phantom behavior,
i.e., the energy density will increase with the scale factor, leading to a super de
Sitter era where the scale factor increases as $a\sim e^{t^2}$. By
contrast, in the $\Lambda$CDM model, the energy density tends to a constant
$\epsilon_\Lambda$ leading to a de Sitter era where the scale factor increases
as $a\sim e^{t}$. Note that the increase of the energy density $\epsilon$
with $a$ in the logotropic model is slow (logarithmic). As a result, there is
no future finite time singularity, i.e., there is no ``big rip'' where 
the energy density and the scale factor become infinite in a finite time
\cite{caldwellprl}. In the logotropic model of type II the energy density and
the scale factor become infinite in infinite time. This is called ``little rip''
\cite{littlerip}.

\subsection{Two-fluid model}
\label{sec_twofluids}

In the model of type II, we have a single dark fluid
with an equation of state $P=P(\rho_{\rm dm})$. Still, the
energy density $\epsilon$ given by Eq. (\ref{mtd2}) is the sum of two terms, a
rest-mass density term
$\rho_{\rm dm}$ which mimics DM and an internal energy term $u(\rho_{\rm dm})$ which mimics
DE. It is interesting to consider a two-fluid model which leads to the
same results as the single dark fluid model, at least for what concerns the
evolution of the homogeneous background. In this two-fluid model, one fluid
corresponds to pressureless DM with an equation of state $P_{\rm dm}=0$ and a
density $\rho_{\rm dm} c^2=\Omega_{\rm dm,0}\epsilon_0/a^3$ determined by the energy
conservation equation for DM, and the other fluid corresponds to DE with an
equation of state $P_{\rm de}(\epsilon_{\rm de})$ and an energy density
$\epsilon_{\rm de}(a)$ determined by the energy
conservation equation for DE. We assume that the two fluids are independent from each other. We can obtain the equation of state of DE yielding
the same results as the one-fluid model by taking
\begin{eqnarray}
P_{\rm de}=P(\rho_{\rm dm}),\qquad \epsilon_{\rm de}=u(\rho_{\rm dm}),
\end{eqnarray}
and eliminating $\rho_{\rm dm}$ from these two relations. In other words, the  equation of state $P_{\rm de}(\epsilon_{\rm de})$ of DE
 in the two-fluid model corresponds to the
relation $P(u)$ in the single fluid model.  We note
that
although
the one and two-fluid models are equivalent for the evolution  of the
homogeneous background, they may differ for what concerns the formation of the
large-scale structures of the universe.

In the two-fluid model associated with a logotrope of type II, the DE
has an affine equation of state \cite{cosmopoly3,jcap}
\begin{eqnarray}
P_{\rm de}=-\epsilon_{\rm de}-A,
\label{twg5}
\end{eqnarray}
which is obtained by eliminating $\rho_{\rm dm}$ between Eqs.
(\ref{eq1bo}) and
(\ref{nrj}), and by identifying $P(u)$ with $P_{\rm de}(\epsilon_{\rm
de})$.  Solving the energy
conservation equation (\ref{econsint}) with the equation of state of DE from
Eq. (\ref{twg5}), we recover Eq. (\ref{nrjb}). However, this two-fluid model
(with $P_{\rm dm}=0$ and $P_{\rm de}=-\epsilon_{\rm de}-A$) does not determine
the value of $A$, contrary to the one-fluid model. This is a huge advantage of
the one-fluid model.

\subsection{Present proportion of DM and DE: Dark magic}
\label{sec_argp}

We now come to a remarkable and very intriguing result. We have seen that the
logotropic model of type II determines a relation between the present fraction
$\Omega_{{\rm dm},0}$ of DM, the present fraction $\Omega_{{\rm de},0}$ of DE
and the present energy density $\epsilon_0$. 
This relation is given by Eq. (\ref{f7}) or, equivalently, by Eq. (\ref{f8})
where we have introduced the notation from Eq. (\ref{impid6}). Then, using
observational results, we have
shown that Eq. (\ref{f8}) can be written in very good approximation as Eq.
(\ref{a2b}). We note that this approximation is valid as long as
\begin{eqnarray}
\frac{\Omega_{{\rm de},0}}{\Omega_{{\rm dm},0}}\ll 10^{124}.
\label{f10}
\end{eqnarray}
This shows that the expression of $A$ is essentially independent from 
the present ratio of DM and DE. Now, we ask ourselves the following question:
What do we get if we assume that Eq. (\ref{a2b}) is {\it exactly} satisfied? 
In that case, we find from Eq. (\ref{f8})
that the ratio between the present proportion of DM and DE is given by the pure
number
\begin{eqnarray}
\frac{\Omega_{{\rm de},0}^{\rm th}}{\Omega_{{\rm dm},0}^{\rm th}}=e=2.71828...
\label{f10a}
\end{eqnarray}
Remarkably, this prediction is in very good agreement with the measured value of
$\Omega_{\rm de,0}^{\rm obs}/\Omega_{\rm dm,0}^{\rm obs}=2.669\pm 0.08$
obtained from $\Omega_{\rm de,0}^{\rm obs}=0.6911\pm 0.0062$  and $\Omega_{\rm
dm,0}^{\rm obs}=0.2589\pm 0.0057$. Using Eq. (\ref{f10a}) and taking into
account the
presence of baryons so that $\Omega_{{\rm dm}}+\Omega_{{\rm de}}+\Omega_{{\rm
b}}=1$, we find that the present proportions 
of DM and DE are
\begin{equation}
\Omega_{{\rm de},0}^{\rm th}=\frac{e}{1+e}(1-\Omega_{{\rm b},0}), \quad \Omega_{{\rm dm},0}^{\rm
th}=\frac{1}{1+e}(1-\Omega_{{\rm b},0}).
\label{f11}
\end{equation}
If we neglect baryonic matter $\Omega_{\rm b,0}=0$ we obtain the pure numbers
$\Omega_{\rm de,0}^{\rm th}=\frac{e}{1+e}=0.731059...$ and
$\Omega_{\rm dm,0}^{\rm
th}=\frac{1}{1+e}=0.268941...$ which give the correct
proportions $70\%$ and $25\%$ of DE and DM \cite{pdu}. If we take 
baryonic matter into account and use the measured value of
$\Omega_{\rm b,0}^{\rm obs}=0.0486\pm 0.0010$, we get
$\Omega_{\rm de,0}^{\rm th}=0.6955\pm 0.0007$ and $\Omega_{\rm dm,0}^{\rm
th}=0.2559\pm
0.0003$ which are very close to the observational values $\Omega_{\rm de,0}^{\rm
obs}=0.6911\pm 0.0062$  and
$\Omega_{\rm dm,0}^{\rm obs}=0.2589\pm
0.0057$ within the error bars. The argument leading to Eq.
(\ref{f10a}) means that the {\it present} DE density $\rho_{\Lambda}$ is
such that
$\rho_{\Lambda}c^2/\ln ({\rho_P}/{\rho_{\Lambda}})$ is equal to the  fundamental
constant $A$ (with infinite precision). This can be viewed as
a 
{\it strong} cosmic coincidence \cite{pdu} giving to our epoch a central
place in the history of the universe.\footnote{Of course, this argument gives
nothing in the framework of the $\Lambda$CDM model since $\rho_{\Lambda}$ is
{\it always} equal to Einstein's cosmological constant $\Lambda/(8\pi G)$. The
argument leading to Eq. (\ref{f10a}) cannot be advocated in the $\Lambda$CDM
model where $\rho_P\rightarrow +\infty$ and $A\rightarrow 0$ because
Eq. (\ref{f8}) degenerates into Eq. (\ref{a2b}).} Until we have an explanation
for this mysterious coincidence  we shall call it {\it dark magic}.

\begin{figure}[!h]
\begin{center}
\includegraphics[clip,scale=0.3]{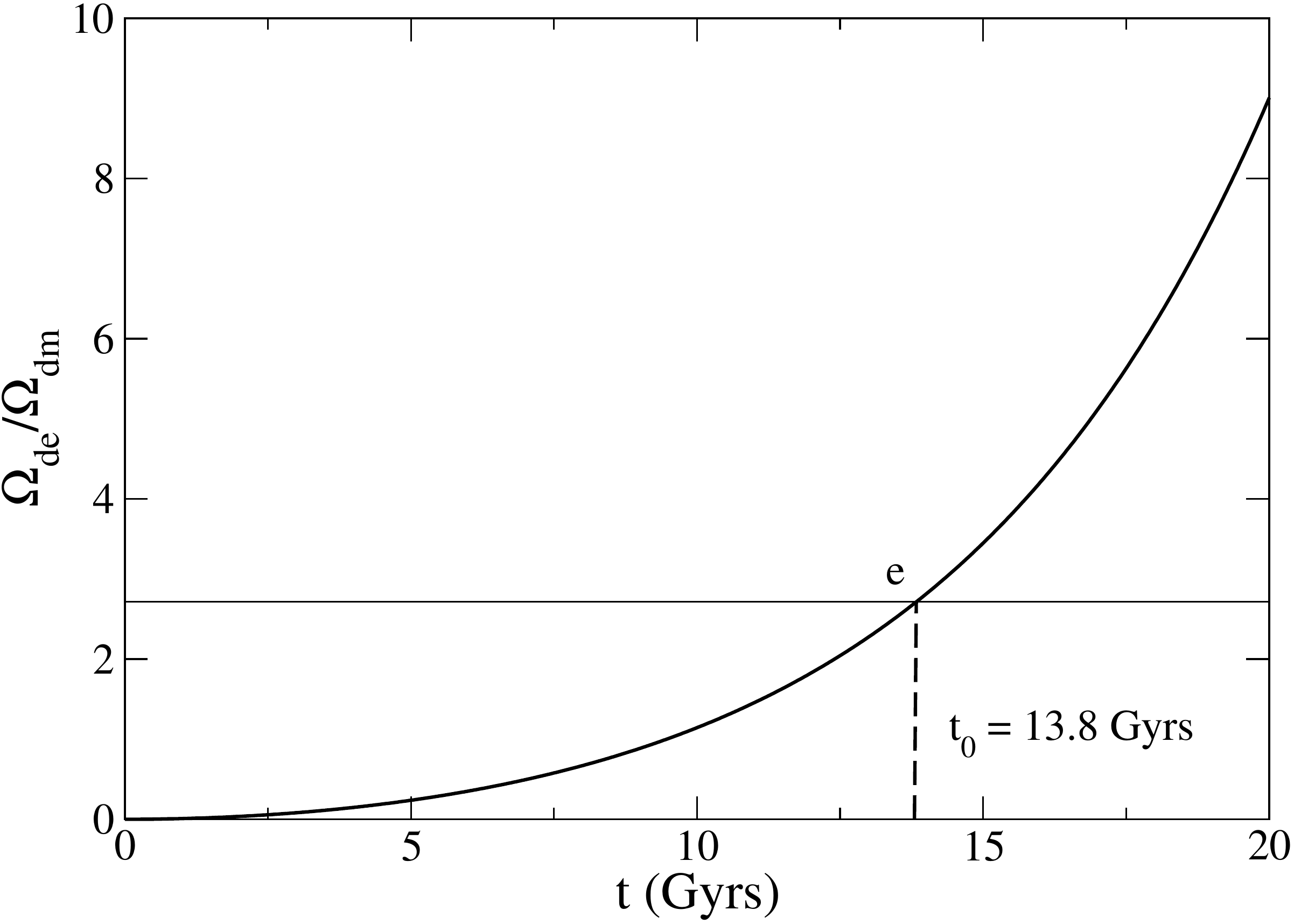}
\caption{Ratio between the proportion of DM and DE as a function of time. Some
arguments based on the logotropic model suggests that this ratio might be equal
to $e=2.71828...$ at the present epoch. This is in agreement with the
observational value $2.669\pm 0.08$.}
\label{ratio}
\end{center}
\end{figure}

{\it Remark:} Using Eqs. (\ref{eq27}) and (\ref{f10a}), we find that the ratio
between the proportion of DM and DE evolves with the scale factor as
\begin{eqnarray}
\frac{\Omega_{\rm de}}{\Omega_{\rm dm}}=ea^3(1+3B\ln a).
\label{f10b}
\end{eqnarray}
It changes algebraically rapidly with the scale factor. This ratio is plotted as
a function of time in Fig. \ref{ratio}. It 
is only at the present epoch ($a=1$) that ${\Omega_{\rm de}}/{\Omega_{\rm
dm}}=e$.

 \section{Logotropic equation of state of type III}
\label{sec_ra}

In this section, we consider a relativistic barotropic fluid  described by an
equation of state of type III where the pressure $P=P(\rho)$ is specified as a
function of the pseudo rest-mass 
density of a complex SF in the Thomas-Fermi (TF) approximation.

\subsection{Klein-Gordon-Einstein equations}
\label{sec_csf}

Let us consider a relativistic  complex SF $\varphi(x^\mu)=\varphi(x,y,z,t)$ which
is a continuous
function of space and time. It can represent the wavefunction of a
relativistic BEC \cite{chavmatos,action}. The total action of the system,  which
is the sum of the
Einstein-Hilbert action of general relativity $+$ the action of the SF, can be
written as
\begin{equation}
S=\int \left (\frac{c^4}{16\pi G}R+ \mathcal{L}\right) \sqrt{-g}\, d^4x,
\label{csf1}
\end{equation}
where $R$ is the Ricci scalar curvature, $\mathcal{L}=\mathcal{L}(\varphi,
\varphi^*,\partial_\mu\varphi,\partial_\mu\varphi^*)$
is the Lagrangian density of the SF,  and
$g={\rm det}(g_{\mu\nu})$ is the
determinant of
the metric tensor.  We consider a
canonical Lagrangian density of the form
\begin{eqnarray}
{\cal L}=\frac{1}{2}g^{\mu\nu}\partial_{\mu}\varphi^*\partial_{\nu}
\varphi-\frac{m^2c^2}{2\hbar^2}|\varphi|^2-V(|\varphi|^2),
\label{csf2}
\end{eqnarray}
where the first term is the kinetic energy, the second term is minus the rest-mass energy
term and the third term is minus the self-interaction energy term.

The least action principle $\delta S=0$ yields the Klein-Gordon-Einstein (KGE) equations
\begin{equation}
\label{csf4b}
\square\varphi+\frac{m^2c^2}{\hbar^2}\varphi+2\frac{dV}{d|\varphi|^2}\varphi=0,
\end{equation}
\begin{equation}
R_{\mu\nu}-\frac{1}{2}g_{\mu\nu}R=\frac{8\pi G}{c^4}T_{\mu\nu},
\label{ak21}
\end{equation}
where
$\square=D_{\mu}\partial^{\mu}=\frac{1}{\sqrt{-g}}\partial_{\mu}(\sqrt{-g}\, g^{
\mu\nu} \partial_{\nu})$ is the d'Alembertian
operator in a curved spacetime, $R_{\mu\nu}$ is the Ricci tensor and $T_{\mu\nu}$ is the
energy-momentum (stress) tensor of the SF given by
\begin{eqnarray}
\label{em2}
T_{\mu\nu}=\frac{1}{2}(\partial_{\mu}\varphi^*\partial_{\nu}\varphi+\partial_{
\nu}\varphi^*\partial_{\mu}\varphi)
-g_{\mu\nu}{\cal L}.
\end{eqnarray}

The conservation of
the energy-momentum tensor,
which results from the invariance of the Lagrangian density under continuous
translations in space and time (Noether theorem), reads
\begin{eqnarray}
\label{lh2r}
D_{\nu}T^{\mu\nu}=0.
\end{eqnarray}
This conservation law can be directly obtained from the Einstein field equations by using the  Bianchi
identities.

The current of charge of the complex SF is given by
\begin{eqnarray}
\label{charge1}
J_{\mu}=-\frac{m}{2i\hbar}
(\varphi^*\partial_\mu\varphi-\varphi\partial_\mu\varphi^*).
\end{eqnarray}
Using the KG equation (\ref{csf4b}), one can show
that 
\begin{eqnarray}
\label{charge2}
D_{\mu}J^{\mu}=0.
\end{eqnarray}
This equation expresses the local conservation of the charge of the SF. The total charge
of the SF is $Q=\frac{1}{mc}\int J^0\sqrt{-g}\, d^3x$, where the elementary charge has been set to unity. In that case, the charge $Q$ is equal to the number $N$ of bosons
provided that antibosons are counted 
negatively \cite{landaulifshitz}. Therefore, Eq. (\ref{charge2})  also
expresses the local conservation of the boson number $N$ or its rest-mass $Nm$. This conservation
law
results via the Noether theorem from the global $U(1)$ symmetry of the
Lagrangian, i.e., from the invariance of the Lagrangian density
under a global phase transformation $\phi\rightarrow \phi e^{-i\theta}$
(rotation) of the
complex SF.

\subsection{Hydrodynamic representation}
\label{sec_db}

We can write the KG equation (\ref{csf4b}) under the form of 
hydrodynamic equations by using the de Broglie
transformation \cite{broglie1927a,broglie1927b,broglie1927c}. To that
purpose, we write the SF as 
\begin{equation}
\varphi=\frac{\hbar}{m}\sqrt{\rho}e^{i S_{\rm tot}/\hbar},
\label{db4}
\end{equation}
where $\rho$ is the pseudo rest-mass density\footnote{We stress that $\rho$ is
{\it not} the rest-mass density $\rho_{\rm dm}=n m$. It is only
in the
nonrelativistic regime $c\rightarrow +\infty$ that $\rho$ coincides with the
rest-mass density.}  defined by
\begin{eqnarray}
\rho=\frac{m^2}{\hbar^2}|\varphi|^2,
\label{db2}
\end{eqnarray} 
and $S_{\rm tot}$ is the action. Substituting Eq. (\ref{db4}) into the Lagrangian density (\ref{csf2}), we obtain
\begin{equation}
{\cal
L}=\frac{1}{2}g^{\mu\nu}\rho\frac{\partial_{\mu} S_{\rm tot}}{m}
\frac{\partial_{\nu}S_{\rm tot}}{m}+\frac{\hbar^2}{8m^2\rho}g^{\mu\nu}\partial_{\mu}\rho\partial_{\nu}
\rho
-\frac{1}{2}\rho c^2-V(\rho).
\label{db5}
\end{equation}

The equations of motion resulting from the least action principle $\delta S=0$ read  \cite{chavmatos,action}
\begin{eqnarray}
D_{\mu}\left ( \rho \frac{\partial^{\mu}S_{\rm tot}}{m}\right )=0,
\label{db8}
\end{eqnarray}
\begin{eqnarray}
\partial_{\mu}S_{\rm tot}\partial^{\mu}S_{\rm tot}=\hbar^2\frac{
\square\sqrt { \rho } } { \sqrt { \rho } } + m^2 c^2 + 2 m^2 V'(\rho).
\label{db9}
\end{eqnarray}
These equations can also be obtained by substituting the de Broglie transformation from
Eq.
(\ref{db4}) into the
KG equation
(\ref{csf4b}),
and by separating the real and the imaginary parts. Equation (\ref{db8}) can be
interpreted as a continuity equation and Eq.
(\ref{db9}) can be interpreted as a  quantum relativistic Hamilton-Jacobi (or
Bernoulli)
equation with a relativistic covariant quantum potential
\begin{eqnarray}
Q_{\rm dB}=\frac{\hbar^2}{2m}\frac{\square\sqrt{\rho}}{\sqrt{\rho}}.
\label{db11}
\end{eqnarray}

The energy-momentum tensor is given, in
the hydrodynamic representation, by 
\begin{eqnarray}
T_{\mu\nu}=\rho\frac{\partial_{\mu}S_{\rm tot}}{m}\frac{\partial_{\nu}S_{\rm tot}}{m}+\frac{\hbar^2}{4m^2\rho}
\partial_{\mu}\rho\partial_{\nu}\rho-g_{\mu\nu}{\cal L}.
\label{em3gen}
\end{eqnarray}
This expression can be obtained from  Eq. (\ref{em2}) by using Eq.
(\ref{db4}). 

The current of charge of a complex SF is given, in
the hydrodynamic representation, by
\begin{eqnarray}
J_{\mu}=-\rho\frac{\partial_{\mu}S_{\rm tot}}{m}.
\label{charge4}
\end{eqnarray}
This expression can be obtained from  Eq. (\ref{charge1}) by using Eq.
(\ref{db4}). We then see that the continuity equation
(\ref{db8}) is
equivalent to Eq. (\ref{charge2}). It
expresses the
conservation of the charge  $Q=-\frac{1}{m^2c}\int \rho\partial^0 S_{\rm tot} \sqrt{-g}  \, d^3x$ of the SF (or, equivalently, the conservation of the boson number $N$).

\subsection{TF approximation}
\label{sec_rtf}

In the classical limit or in the TF approximation ($\hbar\rightarrow 0$), the
Lagrangian from Eq.
(\ref{db5}) reduces to
\begin{eqnarray}
{\cal
L}=\frac{1}{2}g^{\mu\nu}\rho\frac{\partial_{\mu}S_{\rm tot}}{m}\frac{\partial_{\nu}S_{\rm tot}}{m}-\frac{1}{2}\rho c^2-V(\rho).
\label{rtf1}
\end{eqnarray}

The least action principle $\delta S=0$
yields the equations of motion
\begin{eqnarray}
D_{\mu}\left ( \rho 
\frac{\partial^{\mu}S_{\rm tot}}{m}\right )=0,
\label{rtf4}
\end{eqnarray}
\begin{eqnarray}
\partial_{\mu}S_{\rm tot}\partial^{\mu}S_{\rm tot}=m^2c^2+2m^2V'(\rho).
\label{rtf5}
\end{eqnarray}
These equations can also be obtained by making the TF approximation in Eq.
(\ref{db9}), i.e., by neglecting the quantum potential $Q_{\rm dB}$. Equation (\ref{rtf4})
can be interpreted as a continuity equation and Eq.
(\ref{rtf5}) can be interpreted as a classical relativistic Hamilton-Jacobi (or
Bernoulli) equation. Note that the continuity equation is not affected by the TF
approximation.

In the TF approximation, the energy-momentum
tensor (\ref{em3gen}) reduces to
\begin{eqnarray}
T_{\mu\nu}=\rho\frac{\partial_{\mu}S_{\rm tot}}{m}\frac{\partial_{\nu}
S_{\rm tot}}{m}-g_{\mu\nu}{\cal L}.
\label{em3}
\end{eqnarray}
We introduce the fluid quadrivelocity
\begin{eqnarray}
u_{\mu}=-\frac{\partial_{\mu}S_{\rm tot}}{\sqrt{m^2c^2+2m^2V'(\rho)}}c,
\label{rtf5b}
\end{eqnarray}
which satisfies the identity $u_{\mu}u^{\mu}=c^2$. Combining  Eqs. (\ref{em3})
and (\ref{rtf5b}), we get
\begin{eqnarray}
T_{\mu\nu}=\rho\left \lbrack 1+\frac{2}{c^2} V'(\rho)\right \rbrack u_{\mu}u_{\nu}-g_{\mu\nu}{\cal
L}.
\label{em3b}
\end{eqnarray}
The   energy-momentum tensor (\ref{em3b}) can be written under the perfect fluid
form
\begin{eqnarray}
T_{\mu\nu}=(\epsilon+P)\frac{u_{\mu}u_{\nu}}{c^2}-P g_{\mu\nu},
\label{em4}
\end{eqnarray}
where $\epsilon$ is the energy density and $P$ is the pressure, provided that we
make the identifications
\begin{eqnarray}
P={\cal L},\qquad \epsilon+P=\rho c^2+2\rho V'(\rho).
\label{em5}
\end{eqnarray}
Therefore, the Lagrangian density plays the role of the pressure of the fluid. 
Combining  Eq. (\ref{rtf1}) with the Hamilton-Jacobi (or Bernoulli) equation
(\ref{rtf5}), we get
\begin{eqnarray}
{\cal L}=\rho V'(\rho)-V(\rho).
\label{em6}
\end{eqnarray}
Therefore, according to Eqs. (\ref{em5}) and (\ref{em6}), the energy density and
the pressure of the SF in the TF approximation
are given by \cite{abrilphas,action}
\begin{eqnarray}
\epsilon=\rho c^2+\rho V'(\rho)+V(\rho),
\label{rtf6}
\end{eqnarray}
\begin{eqnarray}
P=\rho V'(\rho)-V(\rho).
\label{rtf7}
\end{eqnarray}
Eliminating
$\rho$ between Eqs. (\ref{rtf6}) and (\ref{rtf7}), we obtain the equation of
state $P(\epsilon)$. On the other hand, Eq. (\ref{rtf7}) can be integrated into
\cite{action}
\begin{eqnarray}
V(\rho)=\rho\int\frac{P(\rho)}{\rho^2}\, d\rho.
\label{etoilebis}
\end{eqnarray}
Equation (\ref{rtf7}) determines $P(\rho)$ as a function of $V(\rho)$. Inversely, Eq.
(\ref{etoilebis}) determines $V(\rho)$ as a function of $P(\rho)$.

In the TF approximation, using Eqs. (\ref{charge4}) and (\ref{rtf5b}), we can write the
current as
\begin{eqnarray}
J_{\mu}=\rho\sqrt{1+\frac{2}{c^2}V'(\rho)}\, u_{\mu}.
\label{charge8}
\end{eqnarray}
The rest-mass density $\rho_{\rm dm}=nm$ (which is equal to the charge density) is such that
\begin{eqnarray}
J_{\mu}=\rho_{\rm dm}  u_{\mu}.
\label{charge6w}
\end{eqnarray}
The continuity equation
(\ref{charge2}) can then be written as
\begin{eqnarray}
D_{\mu}(\rho_{\rm dm} u^{\mu})=0.
\label{charge7}
\end{eqnarray}
Comparing Eq. (\ref{charge8}) with Eq. (\ref{charge6w}), we
find that the
rest-mass density of the SF is given
by
\begin{eqnarray}
\rho_{\rm dm}=\rho\sqrt{1+\frac{2}{c^2}V'(\rho)}.
\label{charge10c}
\end{eqnarray}
In general, $\rho_{\rm dm}\neq \rho$ except when $V$ is
constant, corresponding to the $\Lambda$CDM model (see below), and in the
nonrelativistic limit $c\rightarrow +\infty$.

{\it Remark:} in the TF approximation, the equation of state parameter $w=P/\epsilon$ is given by 
\begin{eqnarray}
w=\frac{\rho V'(\rho)-V(\rho)}{\rho c^2+\rho V'(\rho)+V(\rho)}.
\label{rtf7ww}
\end{eqnarray}
For a barotropic equation of state of type III, the universe
exhibits a normal
behavior ($w>-1$) when $1+(2/c^2)V'(\rho)>0$ and a phantom behavior ($w<-1$) 
when $1+(2/c^2)V'(\rho)<0$. In the latter case, the Lagrangian of the SF
involves a negative kinetic  term. The SF has either a normal
behavior (positive kinetic term) or a phantom behavior (negative kinetic term)
but it cannot pass from a normal to a phantom regime. Therefore, a barotropic
equation of state of type III cannot describe the transition from a normal to
a phantom behavior. Here, we only consider 
the normal behavior where $1+(2/c^2)V'(\rho)>0$.

\subsection{Spatially homogeneous SF}
\label{sec_hsf}

For a spatially homogeneous SF in an expanding universe with a Lagrangian
\begin{equation}
L=\frac{1}{2c^2}\left |\frac{d\varphi}{d
t}\right|^2-\frac{m^2c^2}{2\hbar^2}|\varphi|^2-V(|\varphi|^2),
\label{laghom}
\end{equation}
the KG
equation
(\ref{csf4b}) becomes 
\begin{eqnarray}
\frac{1}{c^2}\frac{d^2\varphi}{dt^2}+\frac{3H}{c^2}\frac{d\varphi}{dt}+\frac
{m^2
c^2}{\hbar^2}\varphi
+2\frac{dV}{d|\varphi|^2}\varphi=0,
\label{hsf1}
\end{eqnarray}
while the Einstein field equations (\ref{ak21}) reduce to the Friedmann
equations 
of Sec. \ref{sec_mtuf}. The energy density and the pressure of the SF are given
by
\begin{equation}
\epsilon=\frac{1}{2c^2}\left |\frac{d\varphi}{d
t}\right|^2+\frac{m^2c^2}{2\hbar^2}|\varphi|^2+V(|\varphi|^2),
\label{hsf2}
\end{equation}
\begin{equation}
P=\frac{1}{2c^2}\left |\frac{d\varphi}{d
t}\right|^2-\frac{m^2c^2}{2\hbar^2}|\varphi|^2-V(|\varphi|^2).
\label{hsf3}
\end{equation}
We can
easily check that the KG equation (\ref{hsf1}) with Eqs.  (\ref{hsf2}) and
(\ref{hsf3}) imply the energy conservation equation (\ref{econs}) (see Appendix
G of \cite{action}). In the
following, we use the hydrodynamic representation of the SF (see
Secs. \ref{sec_db} and \ref{sec_rtf}). The total energy  of the SF (including
its rest
mass  energy $mc^2$) is
\begin{eqnarray}
E_{\rm tot}(t)= -\frac{d S_{\rm tot}}{d t}.
\label{ge3}
\end{eqnarray}
For a spatially homogeneous SF, the continuity equation (\ref{db8}) expressing
the conservation of the charge of the SF can be written as \cite{abrilphas}
\begin{eqnarray}
\frac{d}{dt}\left (\rho \frac{E_{\rm tot}}{mc^2} a^3\right )=0.
\label{ge3a}
\end{eqnarray}
It can be
integrated into 
\begin{eqnarray}
\rho\frac{E_{\rm tot}}{mc^2}=\frac{Qm}{a^3},
\label{ge4}
\end{eqnarray}
where $Q$ is the charge of the SF. The rest-mass density $\rho_{\rm dm}$ of a
spatially homogeneous SF is given by
\cite{action}
\begin{eqnarray}
\rho_{\rm dm}=\rho\frac{E_{\rm
tot}}{mc^2}. 
\label{rmd1}
\end{eqnarray}
It is equal to $\rho_{\rm dm}=J_0/c$, where $J_0=-\rho \partial_0S_{\rm tot}/m$
is the time component of the current of charge. This formula is only valid
for a spatially homogeneous SF. Comparing 
Eqs. (\ref{ge4}) and (\ref{rmd1}), we get 
\begin{eqnarray}
\rho_{\rm dm}=\frac{Qm}{a^3}.
\label{rmd2}
\end{eqnarray}
The rest-mass density (or the charge density) of the SF decreases as
$a^{-3}$. This  expresses the conservation
of the charge of the SF or, equivalently, the conservation of the boson
number.\footnote{Inversely,
Eq. (\ref{rmd1}) can be directly obtained from Eq.  (\ref{ge4}) by using Eq.
(\ref{rmd2}).} As in Sec. \ref{sec:cosm}, the rest-mass density of the SF may
be interpreted as DM. Identifying Eq. (\ref{rmd2}) with Eq. (\ref{eq15b}) we
obtain
\begin{eqnarray}
Qmc^2=\rho_{\rm dm,0}c^2=\Omega_{\rm dm,0}\epsilon_0.
\label{log4q}
\end{eqnarray} 
Therefore, the constant $Qm$ (charge) is equal to the
present
density $\rho_{{\rm dm},0}$ of DM  (rest-mass density).

On the other hand, in the TF approximation,  the Hamilton-Jacobi (or Bernoulli)
equation from Eq. (\ref{rtf5}) becomes
\cite{abrilphas}
\begin{eqnarray}
E_{\rm tot}^2=m^2c^4+2m^2c^2V'(\rho).
\label{ge3b}
\end{eqnarray}
It can be
rewritten as
\begin{eqnarray}
E_{\rm tot}=mc^2\sqrt{1+\frac{2}{c^2} V'(\rho)}. 
\label{ge5}
\end{eqnarray}
Combining Eqs. (\ref{ge4})  and (\ref{ge5}), we obtain
\begin{eqnarray}
\rho c^2 \sqrt{1+\frac{2}{c^2}V'(\rho)}=\frac{\Omega_{\rm dm,0}\epsilon_0}{a^3}.
\label{ge6}
\end{eqnarray}
This relation can also be obtained from Eqs. (\ref{charge10c}) and
(\ref{eq16}).

The relation between the energy density and the scale factor is
given by Eq. (\ref{rtf6}) with Eq. (\ref{ge6}). We can then solve
the Friedmann equation (\ref{ak22}) with $\epsilon(a)$ to obtain the temporal evolution of the
scale factor $a(t)$.

\subsection{Logarithmic potential and logotropic equation of state}
\label{sec_log}

We now consider a relativistic complex SF with a
logarithmic potential of the form \cite{logosf}
\begin{equation}
V(|\varphi|^2)=-A\ln \left
(\frac{m^2|\varphi|^2}{\hbar^2\rho_P}\right )-A.
\label{log0}
\end{equation}
The corresponding KG equation reads
\begin{equation}
\Box\varphi+\frac{m^2c^2}{\hbar^2}\varphi-\frac{2A}{|\varphi|^2}\varphi=0.
\label{lwe3}
\end{equation}
This is called the logotropic KG equation. As detailed in \cite{logosf}, we
argue that the potential (\ref{log0}) is not a specific attribute of the SF
(such as its
mass $m$ or self-interaction constant $\lambda$) but that it is an intrinsic
property of the wave equation (\ref{lwe3}).  In other words, we argue that the
logarithmic term involving the fundamental constant $A$ is always present in the
wave equation (\ref{lwe3}) even if, in many situations, it can be neglected
leading to
the ordinary KG equation (corresponding to $A=0$). In this sense, the nonlinear
wave equation
(\ref{lwe3}) is more
fundamental than the linear KG equation.

Using the hydrodynamic
variables introduced in Sec. \ref{sec_db}, the SF potential can be written as
\begin{eqnarray}
V(\rho)=-A\ln\left (\frac{\rho}{\rho_P}\right )-A.
\label{log1}
\end{eqnarray}
In the TF approximation, using Eqs. (\ref{rtf7}) and (\ref{log1}), we find that
the pressure is given
by the logotropic equation of state of type III
\cite{logosf}\footnote{Inversely, we
could
start from the equation of state  (\ref{log2}) and integrate Eq.
(\ref{etoilebis}) to
obtain the
potential $V(\rho)$.}  
\begin{eqnarray}
P=A\ln\left (\frac{\rho}{\rho_P}\right ).
\label{log2}
\end{eqnarray}
On the other hand, using Eqs. (\ref{rtf6}) and (\ref{log1}), we obtain the
energy density
\begin{eqnarray}
\epsilon=\rho c^2-A\ln\left (\frac{\rho}{\rho_P}\right )-2A.
\label{log4}
\end{eqnarray} 
Eliminating $\rho$ between Eqs. (\ref{log2}) and (\ref{log4}), the equation of
state $P(\epsilon)$ is
given in the reversed form $\epsilon(P)$ by
\begin{eqnarray}
\epsilon=e^{P/A} \rho_P c^2-P-2A.
\label{rev}
\end{eqnarray}
This is the equation of state of type I corresponding to the logotropic model of
type III \cite{action}. 

For a spatially homogeneous SF in an expanding universe,
the pseudo
rest-mass density $\rho$ 
evolves according to [see Eq. (\ref{ge6})]
\begin{eqnarray}
\rho c^2 \sqrt{1-\frac{2A}{\rho c^2}}=\frac{\Omega_{\rm dm,0}\epsilon_0}{a^3}.
\label{log3}
\end{eqnarray}
Equation (\ref{log3}) is a second degree equation for $\rho$ which can be 
solved explicitly to give
\begin{eqnarray}
\rho c^2=A+\sqrt{A^2+\frac{(\Omega_{\rm dm,0}\epsilon_0)^2}{a^6}}.
\label{log3b}
\end{eqnarray}
Substituting Eq. (\ref{log3b}) into Eq. (\ref{log4}), the energy density can be
expressed in terms of the scale factor as
\begin{eqnarray}
\epsilon=-A+\sqrt{A^2+\frac{(\Omega_{\rm dm,0}\epsilon_0)^2}{a^6}}\nonumber\\
-A\ln\left
\lbrack \frac{A+\sqrt{A^2+\frac{(\Omega_{\rm
dm,0}\epsilon_0)^2}{a^6}}}{\rho_Pc^2}\right\rbrack.
\label{impid4b}
\end{eqnarray}

\subsection{The value of $A$}
\label{sec_impid}

Applying Eq. (\ref{impid4b}) at the present time ($a=1$) and subtracting the
present contribution $\Omega_{\rm dm,0}\epsilon_0$ of DM we obtain
\begin{eqnarray}
\Omega_{\rm de,0}\epsilon_0=-A+\sqrt{A^2+(\Omega_{\rm dm,0}\epsilon_0)^2}\nonumber\\
-A\ln\left
\lbrack \frac{A+\sqrt{A^2+(\Omega_{\rm
dm,0}\epsilon_0)^2}}{\rho_Pc^2}\right\rbrack-\Omega_{\rm dm,0}\epsilon_0.
\label{impid4bc}
\end{eqnarray} 
Assuming that  $A$ is a universal constant, this equation gives a
relation between $\Omega_{\rm dm,0}$, $\Omega_{\rm de,0}$ and $\epsilon_0$. Inversely, we can use
Eq. (\ref{impid4bc}) and the measured values of $\epsilon_0$, $\Omega_{\rm
dm,0}$
and $\Omega_{\rm de,0}$ to determine the constants of our model. Therefore,
there is no free (undetermined) parameter in our model.

Introducing the notation from Eq. (\ref{impid6}) and writing 
\begin{eqnarray}
A=B\rho_\Lambda c^2,
\label{impid5}
\end{eqnarray}
where $B$ is a dimensionless constant, Eq.
(\ref{impid4bc})  can be rewritten as 
\begin{eqnarray}
1+\frac{\Omega_{\rm
dm,0}}{\Omega_{\rm
de,0}}
=-B+\sqrt{B^2+\left
(\frac{\Omega_{\rm
dm,0}}{\Omega_{\rm
de,0}}\right )^2}\nonumber\\
-B\ln\left\lbrack B+\sqrt{B^2+\left
(\frac{\Omega_{\rm
dm,0}}{\Omega_{\rm
de,0}}\right )^2}\right \rbrack
+B\ln\left(\frac{\rho_P}{\rho_{\Lambda}}\right ).
\label{impid10}
\end{eqnarray}
Using the fact that $B\ll
1$ which can be checked {\it a posteriori}, the foregoing equation reduces to
\begin{eqnarray}
B=\frac{1}{\ln\left(\frac{\rho_P}{\rho_{\Lambda}}\right )-1-\ln\left
(\frac{\Omega_{\rm
dm,0}}{\Omega_{\rm
de,0}}\right )}.
\label{b1}
\end{eqnarray}
We can then redo the discussion from Sec. \ref{sec:cosm}. If we use the
measured values of $\Omega_{\rm
dm,0}$ and  $\Omega_{\rm de,0}$, we find that $A$ and $B$ are given in very
good approximation by Eqs. (\ref{intro10}) and (\ref{intro13}). These results
are largely independent from the ratio $\Omega_{\rm
dm,0}/\Omega_{\rm de,0}$. Inversely, if we impose that $A$ and $B$ are {\it
exactly} given by Eqs. (\ref{intro10}) and (\ref{intro13}), we find
that the ratio  $\Omega_{\rm
dm,0}/\Omega_{\rm de,0}$ is given by Eq. (\ref{f10a}) in very good agreement
with the observed value.
 
 {\it Remark:} Combining Eqs. (\ref{impid6}) and
(\ref{f10a}), we find that
the charge
$Qmc^2=\Omega_{\rm dm,0}\epsilon_0$ 
of the SF [see Eq. (\ref{log4q})] is given by
\begin{eqnarray}
Qmc^2=\frac{\rho_{\Lambda}}{e}. 
\label{cha}
\end{eqnarray}

\subsection{Evolution of the universe}
\label{sec_argev}

Using Eqs. (\ref{log3b}) and (\ref{impid5}), the pressure $P$ and the energy
density $\epsilon$ can be expressed in terms of the scale factor as
\begin{equation}
P = B\Omega_{{\rm de},0}\epsilon_0 \ln\left\lbrack B+\sqrt{B^2+\left
(\frac{\Omega_{\rm
dm,0}}{\Omega_{\rm
de,0}a^3}\right )^2}\right \rbrack
-\Omega_{{\rm de},0}\epsilon_0,
\label{eq1cobc}
\end{equation}
\begin{eqnarray}
\frac{H^2}{H_0^2}=\frac{\epsilon}{\epsilon_0} = -B\Omega_{{\rm de},0} +\sqrt{(B\Omega_{{\rm de},0})^2+\left
(\frac{\Omega_{\rm
dm,0}}{a^3}\right )^2}\nonumber\\
-B\Omega_{{\rm de},0} \ln\left\lbrack B+\sqrt{B^2+\left
(\frac{\Omega_{\rm
dm,0}}{\Omega_{\rm
de,0}a^3}\right )^2}\right \rbrack
+\Omega_{{\rm de},0}.\qquad
\label{eq27b}
\end{eqnarray}
In Eq. (\ref{eq27b}), we have combined the relation $\epsilon(a)$ with the
Friedmann equation
(\ref{ak22}) to obtain a  differential equation determining the temporal
evolution of the scale factor. If we take into account the presence of baryons,
we must add a term $\epsilon_{\rm b}=\Omega_{\rm b,0}\epsilon_0/{a^3}$ in the
energy density. The   $\Lambda$CDM model is recovered for $B=0$ corresponding to
the limit $\hbar\rightarrow 0$ or $\rho_P\rightarrow +\infty$. In that case, the
self-interaction potential is constant ($V=\epsilon_{\Lambda}$)\footnote{In the
complex SF representation of the $\Lambda$CDM model viewed as a UDM model (see
Appendix E of \cite{logosf}), the
total potential is
\begin{equation}
V_{\rm tot}=\frac{m^2c^2}{2\hbar^2}|\varphi|^2+\epsilon_{\Lambda}.
\end{equation}
The constant energy density $\epsilon_{\Lambda}$ does not appear explicitly in
the KG equation which only involves the gradient of the potential. However, it
appears in the energy density and in the pressure. In the TF approximation,
using Eqs. (\ref{rtf6}), (\ref{rtf7}) and (\ref{charge10c}) with
$V=\epsilon_{\Lambda}$ we get $\epsilon=\rho c^2+\epsilon_{\Lambda}$ with
$\rho=\rho_{\rm dm}$ and $P=-\epsilon_{\Lambda}$.} and
Eqs. (\ref{eq1cobc})
and (\ref{eq27b}) return Eqs.
(\ref{intro1}), (\ref{intro2}) and (\ref{intro8}). The
evolution of the universe in the logotropic model of type III has been discussed
in detail in \cite{logosf}. Below, we just recall the main results of this
study.

\begin{figure}[!h]
\begin{center}
\includegraphics[clip,scale=0.3]{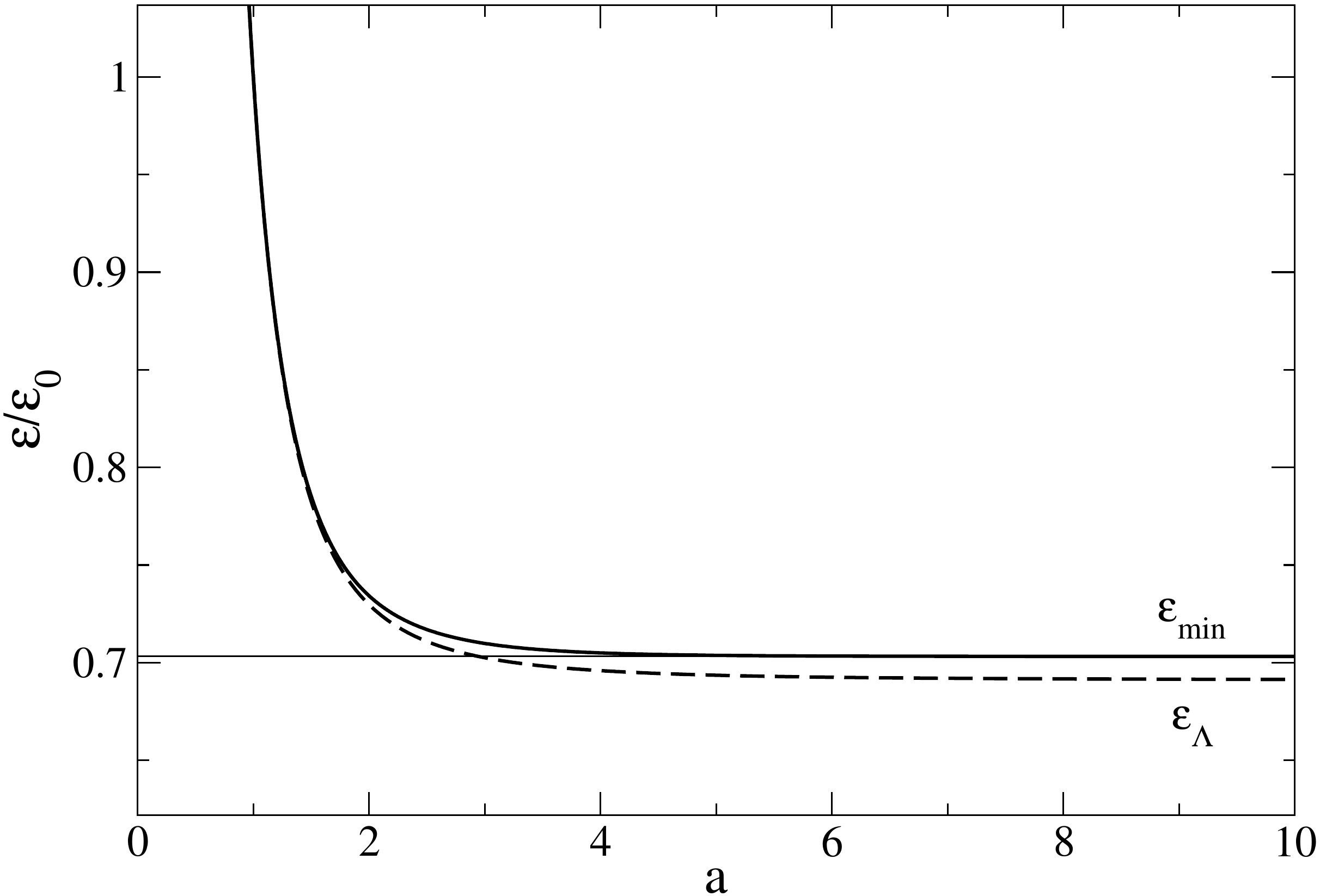}
\caption{Normalized energy density $\epsilon/\epsilon_0$ as a function of the
scale factor $a$ for the logotropic model of type III. It is compared with the
$\Lambda$CDM model. The two curves are indistinguishable at the present time
but their  asymptotes differ by a factor $1.02$.}
\label{aepsT3}
\end{center}
\end{figure}

Starting from $+\infty$, the energy density $\epsilon$ decrease monotonically
and tend to a constant value when $a\rightarrow +\infty$. The pseudo rest-mass
density has a similar behavior. At early times, the pressure is negligible with
respect to the energy density and the LDF is equivalent to a pressureless fluid
like in the CDM model. The energy density decreases as  $\epsilon\sim \rho
c^2\sim \Omega_{\rm dm,0}\epsilon_0/a^3$ and the scale factor increases
algebraically as $a\propto t^{2/3}$ (Einstein-de Sitter). This leads  to a
decelerated expansion of the universe.  At later times, the negative pressure of
the LDF becomes efficient and explains the acceleration of the
universe that we observe today. Ultimately, DE dominates over
DM and the energy density tends 
to a constant $\epsilon_{\rm
min}=\epsilon_{\Lambda}[1-B\ln(2B)]=1.02\, \epsilon_{\Lambda}$. Similarly,
the pseudo rest-mass density tends to $\rho_{\rm min}=2B\rho_{\Lambda}$. The
scale factor has a de Sitter behavior $a\propto {\rm exp}(\sqrt{8\pi
G\epsilon_{\rm min}/3c^2}\, t)$ with a quantum-modified cosmological constant
(i.e. it depends on $B$ hence on $\rho_P$ or $\hbar$). 
The function $\epsilon/\epsilon_0(a)$ is plotted in
Fig. \ref{aepsT3}. 

As the universe expands, the pressure decreases monotonically from $+\infty$ to a minimum value $P_{\rm min}=-\epsilon_{\rm min}$. The pressure $P$ is positive when $\rho>\rho_P$ and negative
when $\rho<\rho_P$. It vanishes at $\rho=\rho_P$. Since the logotropic model is a unification of DM and DE, it is not expected to be valid in the early universe. Therefore, the pressure is always negative in the regime of interest ($\rho\ll \rho_P$) where the logotropic model is valid.

\subsection{Rest-mass density (DM) and internal energy (DE)}
\label{sec_rmd}

The energy density can be written as
\begin{eqnarray}
\epsilon=\rho_{\rm dm} c^2+u=\epsilon_{\rm dm}+\epsilon_{\rm de},
\label{rmd4}
\end{eqnarray}
where the first term is the rest-mass energy and the second term is the
internal energy. As explained in Sec. \ref{sec:cosm}, the rest-mass density
$\rho_{\rm dm}$ represents DM and the internal energy $u$ represents DE. From
Eqs. (\ref{eq16}), (\ref{rtf6}) and (\ref{charge10c}), we
have
\begin{eqnarray}
\epsilon_{\rm dm}=\rho_{\rm dm}c^2=\frac{\Omega_{\rm dm,0}\epsilon_0}{a^3}=\rho c^2 \sqrt{1+\frac{2}{c^2}V'(\rho)},
\label{rmd4b}
\end{eqnarray}
\begin{eqnarray}
\epsilon_{\rm de}&=&u=\epsilon-\epsilon_{\rm dm}\nonumber\\
&=&\rho c^2+V(\rho)+\rho V'(\rho)-\rho c^2\sqrt{1+\frac{2}{c^2}V'(\rho)}.\qquad
\label{rmd6}
\end{eqnarray}
From these equations we  can obtain $u=u(\rho_{\rm dm})$ and $P=P(\rho_{\rm dm})$.

For the LDF,  the rest-mass density $\rho_{\rm dm}$ (DM) is related to the pseudo rest-mass density $\rho$ by
\begin{equation}
\rho_{\rm dm}=\rho\sqrt{1-\frac{2A}{\rho c^2}}.
\label{ie1}
\end{equation} 
This equation can be inverted to give
\begin{equation}
\rho=\frac{A}{c^2}+\sqrt{\frac{A^2}{c^4}+\rho_{\rm dm}^2}.
\label{ie2}
\end{equation} 
On the other hand, the internal energy (DE) is given by
\begin{equation}
\epsilon_{\rm de}=u=\rho c^2-A\ln\left (\frac{\rho}{\rho_P}\right )-2A-\rho c^2\sqrt{1-\frac{2A}{\rho c^2}}.
\label{log4b}
\end{equation} 
Substituting Eq. (\ref{ie2}) into Eq. (\ref{log4b}) we get
\begin{eqnarray}
\epsilon_{\rm de}=u=&-&A+\sqrt{A^2+\rho_{\rm dm}^2c^4}-\rho_{\rm dm}
c^2\nonumber\\
&-&A\ln\left\lbrack
\frac{A}{\rho_Pc^2}+\sqrt{\frac{A^2}{\rho_P^2c^4}+\frac{\rho_{\rm
dm}^2}{\rho_P^2}}
\right\rbrack,
\label{ie3}
\end{eqnarray}
which determines $u(\rho_{\rm dm})$. On the other hand, according to Eqs.
(\ref{log2}) and (\ref{ie2}), we obtain the equation
of state $P(\rho_{\rm dm})$ of the SF in terms of the rest-mass density as
\begin{eqnarray}
P=A\ln\left\lbrack
\frac{A}{\rho_Pc^2}+\sqrt{\frac{A^2}{\rho_P^2c^4}+\frac{\rho_{\rm
dm}^2}{\rho_P^2}}
\right\rbrack.
\label{ie4}
\end{eqnarray} 
This is the equation of state of type II corresponding to the logotropic model
of type III \cite{action}.

\subsection{Two-fluid model}
\label{sec_rmdtf}

In the two-fluid model (see Sec. \ref{sec_twofluids}), the DE
has an equation of state $P_{\rm de}(\epsilon_{\rm
de})$ which is obtained by eliminating $\rho$ between Eqs. (\ref{rtf7})
and
(\ref{rmd6}), and by identifying $P(u)$ with $P_{\rm de}(\epsilon_{\rm
de})$. For the logotropic model, it
can be written in inverse form
as\footnote{This relation can be
obtained from Eqs. (\ref{log2}) and (\ref{log4b}). It can also be
obtained by solving Eq. (\ref{ie4}) to get $\rho_{\rm dm}(P)$
and by using Eqs. (\ref{rev}) and (\ref{rmd4}).}
\begin{eqnarray}
\epsilon_{\rm de}=\rho_P c^2 e^{P_{\rm de}/A}-P_{\rm de}-2A\nonumber\\
-\rho_P
c^2 \sqrt{e^{2P_{\rm de}/A}-\frac{2A}{\rho_P c^2}e^{P_{\rm de}/A}}.
\label{log6ju}
\end{eqnarray}

\section{Logotropic DM halos}
\label{sec_pldm}

In this section, we describe the structure of logotropic DM halos and determine their universal surface density.
We use a nonrelativistic approach that is appropriate to DM halos. For 
the sake of generality, we consider the logotropic model of type III which is
based on a complex SF theory. However, after giving general results, we shall 
make the TF approximation.  In that case, the models of type I, II and III are
equivalent in the nonrelativistic limit so our results are valid for all these
models.  In the nonrelativistic limit, the mass density is equal to 
$\rho=\rho_{\rm dm}=\epsilon/c^2$ and the SF potential $V$ coincides with the
internal energy $u$ \cite{action}.

 \subsection{Logotropic GP equation} 
\label{sec_fws}

Basically, a complex SF is  governed by the KGE equations
(\ref{csf4b}) and (\ref{ak21}). In the nonrelativistic limit $c\rightarrow
+\infty$, making the Klein
transformation \cite{klein1,klein2,chavmatos}
\begin{eqnarray}
\label{klein}
\varphi({\bf r},t)=\frac{\hbar}{m}e^{-imc^2t/\hbar}\psi({\bf r},t),
\end{eqnarray}
the KGE equations (\ref{csf4b}) and (\ref{ak21}) reduce
to the GPP equations\footnote{We consider here a static
background ($a=1$) since we will discuss these equations in the context of DM
halos where the expansion of the universe can be neglected (see
\cite{aacosmo,jeansuniverse,jeansMR} for generalizations).}
\begin{eqnarray}
i\hbar\frac{\partial\psi}{\partial
t}=-\frac{\hbar^2}{2 m }\Delta\psi+m\Phi \psi
+m\frac{dV}{d|\psi|^2}\psi,
\label{lwe4}
\end{eqnarray}
\begin{eqnarray}
\Delta\Phi=4\pi G|\psi|^2, 
\label{lwe5}
\end{eqnarray}
where $\psi$ is the wavefunction such that
$\rho=|\psi|^2=(m/\hbar)^2|\varphi|^2$. We refer to
\cite{playa,chavmatos} for a detailed derivation of these equations.
For the logarithmic potential
\begin{eqnarray}
V(|\psi|^2)=-A\ln\left (\frac{|\psi|^2}{\rho_P}\right )-A,
\end{eqnarray}
corresponding to Eq. (\ref{log1}), we obtain 
the logotropic GP equation \cite{epjp}
\begin{eqnarray}
i\hbar \frac{\partial\psi}{\partial
t}=-\frac{\hbar^2}{2m}\Delta\psi+m\Phi\psi-\frac{Am}{|\psi|^2}\psi.
\label{lwe6}
\end{eqnarray} 
For $A=0$, corresponding to a constant potential (see Sec. \ref{sec_ra}), we
recover the
Schr\"odinger-Poisson equations of the fuzzy dark matter (FDM)
model \cite{hu,hui}.

\subsection{Madelung transformation} 
\label{sec_mad}

Writing the wave function as
\begin{equation}
\label{mad1}
\psi({\bf r},t)=\sqrt{{\rho({\bf r},t)}} e^{iS({\bf r},t)/\hbar},
\end{equation}
where $S({\bf r},t)$ is the action,
and making the Madelung \cite{madelung} transformation
\begin{equation}
\label{mad2}
{\bf u}=\frac{\nabla S}{m},
\end{equation}
where ${\bf u}({\bf r},t)$ is the velocity field, the GPP
equations (\ref{lwe4}) and (\ref{lwe5}) can be written under the form of
hydrodynamic equations as
\begin{equation}
\label{mad3}
\frac{\partial\rho}{\partial t}+\nabla\cdot (\rho {\bf u})=0,
\end{equation}
\begin{equation}
\label{mad4}
\frac{\partial {\bf u}}{\partial t}+({\bf u}\cdot
\nabla){\bf
u}=-\frac{1}{m}\nabla
Q_B-\frac{1}{\rho}\nabla P-\nabla\Phi,
\end{equation}
\begin{equation}
\label{mad5}
\Delta\Phi=4\pi G \rho,
\end{equation}
where
\begin{equation}
\label{mad6}
Q_B=-\frac{\hbar^2}{2m}\frac{\Delta
\sqrt{\rho}}{\sqrt{\rho}}=-\frac{\hbar^2}{4m}\left\lbrack
\frac{\Delta\rho}{\rho}-\frac{1}{2}\frac{(\nabla\rho)^2}{\rho^2}\right\rbrack
\end{equation}
is the Bohm quantum potential taking into account the Heisenberg uncertainty
principle and $P(\rho)$ is the pressure determined by the potential according to
Eq. (\ref{rtf7}). For the
logarithmic potential (\ref{log1}), we obtain the logotropic equation
of state (\ref{log2}).

\subsection{Condition of  hydrostatic
equilibrium} 
\label{sec_madeq}

In
this section, we make the TF approximation which
amounts to
neglecting the quantum potential.\footnote{The effect of the quantum potential
is discussed in \cite{logosf}. In that case, the  DM halos exhibit an
additional quantum core corresponding to a noninteracting self-gravitating BEC
(soliton) like in the FDM model \cite{ch2,ch3}.} In
that case, the
equilibrium state of a DM halo results from the balance between
the gravitational attraction and the repulsion due to the pressure
force. It is described by the classical
equation of
hydrostatic equilibrium
\begin{eqnarray}
\nabla P+\rho\nabla\Phi={\bf 0}
\label{diff4}
\end{eqnarray}
coupled to the Poisson equation
\begin{eqnarray}
\Delta\Phi=4\pi G\rho.
\label{diff5}
\end{eqnarray} 
These equations  can be combined into a single differential equation
\begin{eqnarray}
-\nabla\cdot \left(\frac{\nabla
P}{\rho}\right )=4\pi G\rho,
\label{diff5b}
\end{eqnarray}
which determines the density
profile of a DM halo. For the logotropic equation of state
(\ref{log2}), it becomes
\begin{eqnarray}
A\Delta\left(\frac{1}{\rho}\right )=4\pi G\rho.
\label{lel1}
\end{eqnarray}
If we define
\begin{equation}
\label{lel2}
\theta=\frac{\rho_0}{\rho},\qquad \xi=\left (\frac{4\pi
G\rho_0^2}{A}\right )^{1/2}r,
\end{equation}
where $\rho_0$ is the central density and $r_0=({A}/{4\pi
G\rho_0^2})^{1/2}$ is the logotropic core radius, we find that Eq.
(\ref{lel1})
reduces to the Lane-Emden equation of index $n=-1$
\cite{chandrabook}:
\begin{equation}
\label{lel3}
\frac{1}{\xi^2}\frac{d}{d\xi}\left (\xi^2
\frac{d\theta}{d\xi}\right )=\frac{1}{\theta},
\end{equation}
with the boundary conditions $\theta=1$ and $\theta'=0$ at $\xi=0$. This
equation has been studied in detail in  \cite{logo,epjp}. There exists an exact
analytical solution $\theta_s=\xi/\sqrt{2}$, corresponding to $\rho_s=(A/8\pi
G)^{1/2}r^{-1}$,
called
the singular logotropic sphere. The regular logotropic density profiles must be
computed numerically. The normalized density profile $\rho/\rho_0(r/r_0)$ is
universal as a consequence of the homology invariance of
the solutions of the Lane-Emden equation. It is
plotted in Fig. 18  of \cite{epjp}. The density
profile of a logotropic DM halo has a core
($\rho\rightarrow {\rm cst}$ when $r\rightarrow 0$) and decreases at large
distances as $\rho\sim
r^{-1}$. More precisely, for $r\rightarrow +\infty$, we have
\begin{eqnarray}
\rho\sim \left (\frac{A}{8\pi G}\right )^{1/2}\frac{1}{r},
\end{eqnarray}
like for the singular logotropic sphere. This profile has an infinite mass
because the density does not decrease sufficiently rapidly with the distance.
This implies
that, in the case of real DM
halos, the logotropic equation of state (\ref{log2}) or the logotropic profile
determined by Eq. (\ref{lel1}) cannot be valid at
infinitely
large distances (corresponding to very low
densities). Actually, the logotropic core is expected to be surrounded by an extended
envelope resulting from a process of violent collisionless relaxation
\cite{lb,wignerPH,logosf}.\footnote{In the quantum logotropic
model based on the logotropic GPP equations (\ref{lwe5}) and (\ref{lwe6}), the
envelope is due to quantum interferences of excited states, like in the FDM
model \cite{ch2,ch3}.}
In the envelope, the density decreases more rapidly than $r^{-1}$, typically 
like $r^{-2}$ corresponding to the isothermal sphere \cite{lb,wignerPH,logosf}
or like $r^{-3}$ similar to the NFW \cite{nfw} and Burkert \cite{burkert}
profiles. In the following, we shall consider the logotropic profile up to a
few halo radii $r_h$ so we do not have to consider the effect of the envelope.

\subsection{Halo mass}
\label{sec_hma}

\begin{figure}[!h]
\begin{center}
\includegraphics[clip,scale=0.3]{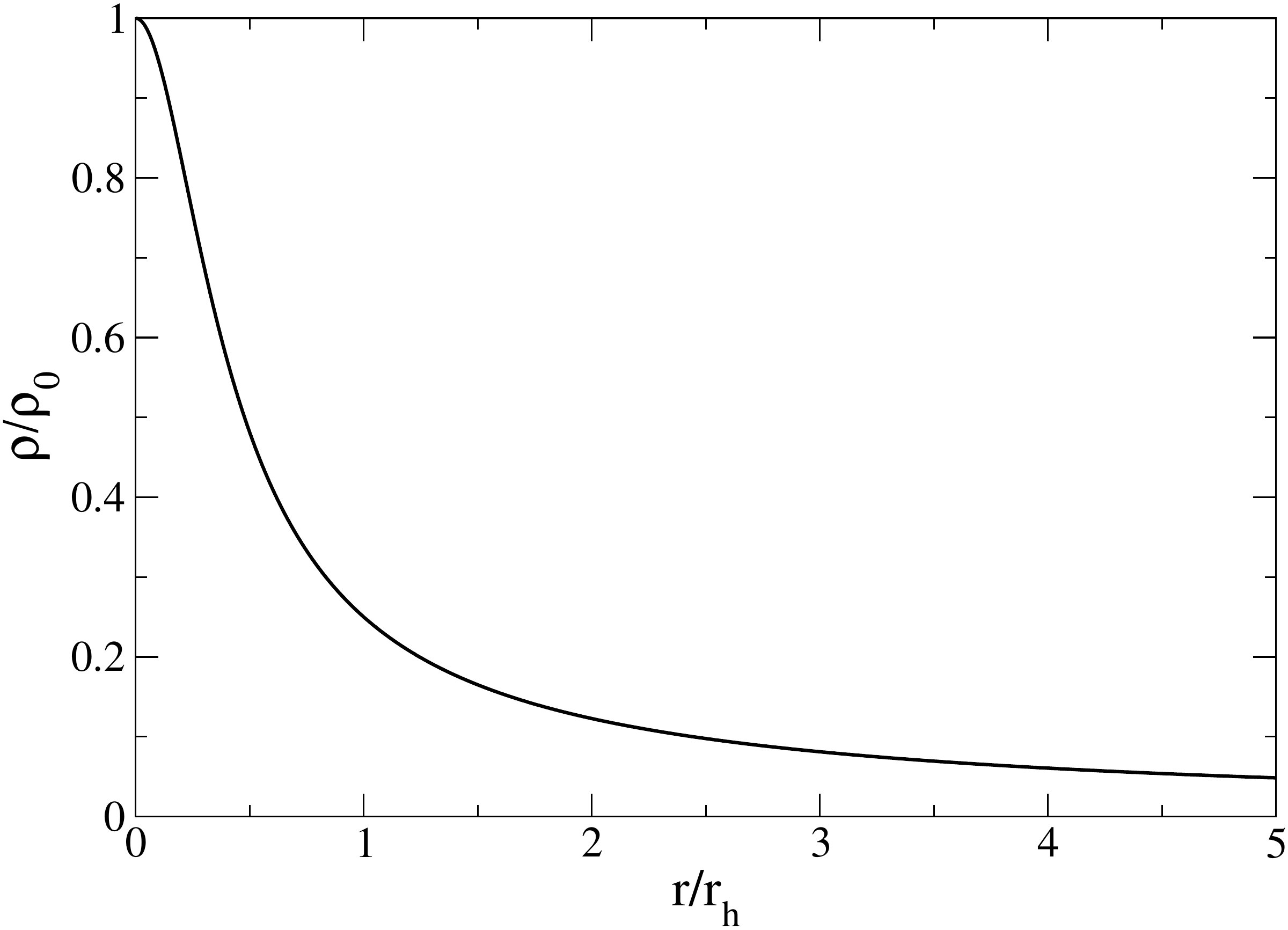}
\caption{Normalized density profile of a logotropic DM halo.}
\label{densLOGO}
\end{center}
\end{figure}

The halo radius $r_h$ is defined as the distance at which the
central density  $\rho_0$  is divided by $4$. For logotropic DM halos, using Eq.
(\ref{lel2}), it is
given by
\begin{eqnarray}
\label{lel4}
r_h=\left (\frac{A}{4\pi
G\rho_0^2}\right )^{1/2}\xi_h,
\end{eqnarray} 
where
$\xi_h$ is determined by the equation
\begin{eqnarray}
\label{lel5}
\theta(\xi_h)=4.
\end{eqnarray} 
The normalized density profile $\rho/\rho_0(r/r_h)$ of
logotropic DM halos is
plotted in Fig.
\ref{densLOGO}. The halo
mass
$M_h$,
which is the
mass $M_h=\int_0^{r_h} \rho(r') 4\pi {r'}^2\, dr'$ contained within the sphere
of radius $r_h$, is given by \cite{epjp} 
\begin{eqnarray}
\label{lel6}
M_h=4\pi\frac{\theta'(\xi_h)}{{\xi_h}}\rho_0 r_h^3.
\end{eqnarray} 
Solving the Lane-Emden equation of index $n=-1$ [see Eq. (\ref{lel3})], we
numerically find \cite{epjp}
\begin{eqnarray}
\xi_h=5.85,\qquad \theta'_h=0.693.
\label{lel7}
\end{eqnarray}
This yields
\begin{eqnarray}
r_h=5.85\, \left (\frac{A}{4\pi G}\right )^{1/2}\frac{1}{\rho_0}
\label{lel8}
\end{eqnarray}
and
\begin{eqnarray}
M_h=1.49\, \rho_0 r_h^3.
\label{lel9}
\end{eqnarray}

\subsection{Constant surface density}
\label{sec_cds}

Eliminating the central density between Eqs. (\ref{lel8}) and (\ref{lel9}), we
obtain the logotropic halo mass-radius relation
\begin{eqnarray}
M_h=8.71 \, \left (\frac{A}{4\pi G}\right
)^{1/2} r_h^2.
\label{lel11}
\end{eqnarray}
Since $M_h\propto r_h^2$ we see that the surface density $\Sigma_0$ is 
constant.\footnote{This is a consequence of the fact that the density of a
logotropic DM halo decreases as $r^{-1}$ at large distances.} This is a very
important property of logotropic DM
halos \cite{epjp}. From Eq. (\ref{lel8}), we get 
\begin{eqnarray}
\Sigma_0=\rho_0 r_h=5.85 \, \left (\frac{A}{4\pi G}\right
)^{1/2}.
\label{lel12}
\end{eqnarray}
Therefore, all the logotropic DM halos have the same surface density, whatever
their size, provided that
$A$ is interpreted as a universal constant. With
the value of
$A/c^2=2.10\times
10^{-26}\, {\rm g}\, {\rm m}^{-3}$ obtained (without free parameter) from the
cosmological
considerations  of Secs.
\ref{sec_mtu}-\ref{sec_ra}  we get
$\Sigma_0^{\rm
th}=133\,
M_{\odot}/{\rm pc}^2$ in very good agreement with the value
$\Sigma_0^{\rm obs}=\rho_0 r_h=141_{-52}^{+83}\, M_{\odot}/{\rm
pc}^2$ obtained from the observations \cite{donato}.   On the other hand, Eq.
(\ref{lel9}) may
 be rewritten as 
\begin{eqnarray}
M_h=1.49\, \Sigma_0 r_h^2=1.49\, \frac{\Sigma_0^3}{\rho_0^2}.
\label{lel13}
\end{eqnarray}
We note that the ratio $M_h/(\Sigma_0 r_h^2)=1.49$ in Eq. (\ref{lel13}) is  in
good
agreement with the ratio $M_h/(\Sigma_0 r_h^2)=1.60$ obtained from the
observational Burkert profile (see Appendix D.4 of \cite{modeldmB}). This is an
additional argument in favor of the logotropic model.

\subsection{Alternative expressions of the universal surface density}
\label{sec_cdsa}

We can write the universal surface density of DM halos
\begin{eqnarray}
\Sigma_0^{\rm th}=\left (\frac{A}{4\pi G}\right )^{1/2}\xi_h=133\,
M_{\odot}/{\rm pc}^2
\label{diff6}
\end{eqnarray} 
in terms of the Einstein cosmological constant $\Lambda$ interpreted as an
effective constant in our model (defined in terms of the present DE density).
Using
$A=B\rho_{\Lambda}c^2$ and
$\rho_{\Lambda}=\Lambda/(8\pi G)$, we get
\begin{eqnarray}
\Sigma_0^{\rm th}=\left (\frac{B}{32}\right
)^{1/2}\frac{\xi_h}{\pi}\frac{c\sqrt{\Lambda}}{G}=0.01955\frac{c\sqrt{
\Lambda}}{G},
\label{diff7}
\end{eqnarray}
where we have used the numerical values of $B=3.53\times 10^{-3}$ and
$\xi_h=5.85$.
Recalling
that $B$ is given by Eq. (\ref{intro13}) with $\rho_P/\rho_{\Lambda}=8\pi
c^5/\hbar G\Lambda$, we also have
\begin{eqnarray}
\Sigma_0^{\rm th}= 0.329  
\frac{\frac{c\sqrt{\Lambda}}{G}}{\sqrt{\ln\left(\frac{8\pi c^5}{\hbar
G\Lambda}\right )}}.
\label{diff8}
\end{eqnarray}
Since $\rho_{\Lambda}$ represents 
the present density of DE, it may be more relevant to express $\Sigma_0^{\rm
th}$ in terms of the present value of the Hubble constant $H_0$. Using
$\Lambda=3\Omega_{\rm de,0}H_0^2$ obtained from Eqs. (\ref{intro3}),
(\ref{ak22}) and (\ref{impid6}), we get
\begin{eqnarray}
\Sigma_0^{\rm th}=0.02815\frac{H_0
c}{G}.
\label{diff7b}
\end{eqnarray}
These identities express the universal surface density of DM halos in terms of
the fundamental constants of physics $G$, $c$,  $\hbar$ and $\Lambda$ (or $A$). We
stress that the prefactors are determined by our model so there is no free parameter. We note that the 
identities from Eqs. (\ref{diff6})-(\ref{diff7b}), which can be checked by
a
direct numerical application, are interesting in themselves 
even in the case where the logotropic model would
turn out to be wrong. Furthermore, as observed in \cite{pdu}, the surface
density of DM halos is of the same order of magnitude as the
surface density of the universe and, more surprisingly, as the surface density
of the electron. As a result, the identities from Eqs.
 (\ref{diff7})-(\ref{diff7b}) allow us to express
the mass of the electron\footnote{We remain here at a very qualitative level
and ignore dimensionless prefactors. The mass scale appearing in Eqs.
(\ref{edd}) and (\ref{wein}) may correspond to the mass of the electron, nucleon
(proton or neutron), pion...} in
terms of the cosmological constant and the other fundamental constants of
physics as \cite{pdu}
\begin{eqnarray}
m_e\sim \left (\frac{\Lambda\hbar^4}{G^2c^2}\right )^{1/6}
\label{edd}
\end{eqnarray}
or as
\begin{eqnarray}
m_e\sim \left (\frac{H_0\hbar^2}{Gc}\right )^{1/3}.
\label{wein}
\end{eqnarray}
This returns the empirical Eddington-Weinberg relation
\cite{eddington,weinbergbook} obtained from different considerations. This
relation 
provides a curious connection between microphysics and macrophysics (i.e.
between atomic physics and cosmology) which is
further discussed in \cite{pdu,oufsuite}. Curiously, the Weinberg 
relation (\ref{wein}) involves the {\it present} value of the Hubble constant
(see the conclusion).

\subsection{The gravitational acceleration}
\label{sec_ga}

We can define an average DM halo surface density by the relation
\begin{eqnarray}
\label{lel14}
\langle\Sigma\rangle=\frac{M_h}{\pi r_h^2}.
\end{eqnarray}
For logotropic DM halos, using Eq. (\ref{lel13}), we find
\begin{eqnarray}
\label{lel15}
\langle\Sigma\rangle_{\rm th}=\frac{M_h}{\pi
r_h^2}=\frac{1.49}{\pi}\Sigma_0^{\rm th}=63.1\,
M_{\odot}/{\rm pc}^2.
\end{eqnarray}
This theoretical value is in good agreement with the
value
$\langle\Sigma\rangle_{\rm
obs}=72_{-27}^{+42}, M_{\odot}/{\rm pc}^2$ obtained from the observations
\cite{gentile}.

The gravitational acceleration at the halo radius is
\begin{eqnarray}
\label{lel16}
g=g(r_h)=\frac{GM_h}{r_h^2}=\pi G \langle\Sigma\rangle.
\end{eqnarray}
For logotropic DM
halos, we find
\begin{equation}
\label{lel17}
g_{\rm th}=\pi G \langle\Sigma\rangle_{\rm
th}=1.49 G \Sigma_0^{\rm
th}=2.76\times
10^{-11}\, {\rm m/s^2}.
\end{equation}
Again, this theoretical value is in good agreement with the measured value
$g_{\rm obs}=\pi G
\langle\Sigma\rangle_{\rm obs}=3.2_{-1.2}^{+1.8}\times 10^{-11}\, {\rm
m/s^2}$ of the gravitational acceleration \cite{gentile}.

The circular velocity at the halo radius  is 
\begin{equation}
\label{lel18}
v_h^2=\frac{GM_h}{r_h}.
\end{equation}
Using Eqs. (\ref{lel15})-(\ref{lel17}), we obtain the relation
\begin{equation}
\label{lel18b}
v_h^4=GgM_h=\pi\langle\Sigma\rangle G^2M_h=1.49\Sigma_0G^2M_h,
\end{equation}
where $g$ and $\Sigma_0$ are universal constants. This relation is connected 
to the Tully-Fisher relation \cite{tf}  which involves the baryon mass
$M_b$
instead of the DM halo mass $M_h$ via the cosmic baryon fraction
$f_b=M_b/M_h\sim 0.17$. This yields $(M_{\rm
b}/v_h^4)^{\rm th}=46.4\,
M_{\odot}{\rm km}^{-4}{\rm s}^4$ which is close to the
observed value 
$(M_{\rm b}/v_h^4)^{\rm obs}=47\pm 6 \, M_{\odot}{\rm km}^{-4}{\rm s}^4$
\cite{mcgaugh}. The Tully-Fisher relation is also a
prediction of the MOND
(modification of Newtonian dynamics) theory \cite{mond}. Using Eqs.
(\ref{diff7b}) and (\ref{lel17}),
we obtain
\begin{eqnarray}
g_{\rm th}=0.0291\sqrt{3\Omega_{\rm de,0}}H_0 c=0.0419\, H_0 c.
\label{diff7c}
\end{eqnarray}
This relation explains why the fundamental constant $a_0=g/f_b$ that appears in
the MOND theory is of order $H_0 c/4=1.65\times
10^{-10}\, {\rm m/s^2}$ (see the Remark in Sec. 3.3. of \cite{pdu} for a more
detailed discussion). Note, however, that our model is completely different from
the MOND
theory.

\section{Conclusion}
\label{sec_ldm}

In this paper, we have first discussed the similarities and the differences
between three types of logotropic models. The logotropic model of type I where
the pressure is proportional to the logarithm of the energy density is
indistinguishable from
the $\Lambda$CDM model, at least for what concerns the evolution of the
homogeneous background. The   logotropic model of type II where the pressure is
proportional to the logarithm of the rest mass density will 
differ from the $\Lambda$CDM model in about $27\, {\rm Gyrs}$ years. At that
moment it will present a phantom behavior in which the energy density increases
with the scale factor (leading to a little rip) while the energy density of the
$\Lambda$CDM model tends
to a constant. Finally, the   logotropic model of type III where the pressure is
proportional to the logarithm of the pseudo rest mass density of a complex SF is
similar to the
$\Lambda$CDM model except that its asymptotic energy density differs from the
asymptotic energy density of the  $\Lambda$CDM model by a factor $1.02$.

We have then emphasized two main predictions of the logotropic model. These predictions were made in our previous papers but we have improved and generalized our argumentation. 

The first prediction is the universality of the surface density of DM halos.
This is a direct consequence of the logotropic equation of state which implies
an asymptotic  density profile of the form $\rho\propto r^{-1}$ or a mass-radius
relation of the form $M_h\propto r_h^2$. The surface density is determined by
the logotropic constant $A$. If this constant is interpreted as a fundamental
constant of physics, it immediately explains the universality of the surface
density of DM halos. Interestingly, we have determined the value of this
constant from cosmological (large scale) considerations without free parameter.
The constant $A$ is
related to the
cosmological density $\rho_{\Lambda}$, defined in our model 
as the present DE density, by Eq. (\ref{intro10}).  This value ensures that the
logotropic model explains the evolution  of the homogeneous background up to the
present time
as well as the $\Lambda$CDM model. Then, using the value of $A$ to determine the
universal surface density of DM halos (small scales) we obtained  Eq.
(\ref{diff6}) which is in very good agreement with the observational
value.
Therefore, the fundamental constant $A$ appearing in the logotropic model is
able to account both for the large scale (cosmological) and the small scale
(astrophysical) properties of our universe.\footnote{By contrast, the
$\Lambda$CDM model works well at large scales but faces a small scale
crisis \cite{crisis}
(see the introduction).} Indeed, it explains both the
acceleration of the universe and the universality of the surface density of DM
halos.
Intriguingly, there also seems to exist a connection between
cosmological, astrophysical and
atomic scales which manifests itself in the commensurability of the surface
density of DM halos (or the surface density of the universe) and the surface
density of the electron \cite{pdu}. This ``coincidence'' may shed a new light on
the
mysterious Eddington-Weinberg relation which relates the mass of the electron to
the cosmological constant or to the present value of the Hubble
constant \cite{pdu,oufsuite}.

The second prediction of the logotropic model is even more intriguing. We have
argued that the present ratio $\Omega_{\rm de,0}/\Omega_{\rm dm,0}$ of DE and
DM is equal to the pure number $e=2.71828...$. This prediction lies in the
error bars of the measured value $\Omega_{\rm de,0}^{\rm
obs}/\Omega_{\rm dm,0}^{\rm obs}=2.669\pm 0.08$.
This result is striking because the proportion of DE and DM changes with time 
so it is only at the present epoch that their ratio is equal to $e$. Indeed, to
obtain
this result, we have assumed that the {\it present} proportion of DE
$\rho_\Lambda$ is related to the fundamental constant $A$ (independent of time)
by the relation from Eq. (\ref{intro10}). This gives to our present epoch a
particular place
in the history of the universe. This coincidence is mysterious and disturbing.
It is almost mystical (dark magic).
It can be viewed as a refined form of the well-known cosmic coincidence
problem \cite{stein1,zws},
namely why $\Omega_{\rm de}/\Omega_{\rm dm}$ is of order one today. If our
result
is confirmed, we have to explain not only why   $\Omega_{\rm de,0}/\Omega_{\rm
dm,0}$ is of order one but, more precisely, why it is equal to $e$. We call
it the strong cosmic coincidence problem. This new
coincidence may  be related in some sense to the coincidence of large numbers
noted by Eddington, Dirac and Weinberg. For example, the Weinberg relation
(\ref{wein})
relates the mass of the electron to the {\it present} value of the Hubble
constant (or equivalently to the present age of the universe). Therefore, it
also gives to our present epoch a particular place in the history of the
universe. To avoid this coincidence, Dirac \cite{dirac} proposed that the
fundamental constants of physics (e.g. the gravitational constant) change with
time so that the relation between
the large numbers that we observe today is always valid. For
example, the Weinberg relation would always be valid if the gravitational
constant changes with time as $G\sim t^{-1}$. Unfortunately, the observations
do not
support Dirac's theory about a time-varying gravitational constant
\cite{dgt,uzan}. As a result, these coincidences and the special place
of
our epoch in the history of the universe remain a mystery. We must either accept
that our epoch plays a particular role in the cosmological evolution (which
would  gives to mankind a central place in the universe like in the old
geocentric theory) or reformulate the laws of physics and cosmology so that what
appears to be a coincidence at the present epoch finds a natural
justification. The relation $\Omega_{\rm de,0}/\Omega_{\rm dm,0}=e$ may
correspond to a fixed point in a more sophisticated theory.

Despite these interesting and intriguing results, there are two main issues with
the logotropic model:

(i) The density profile of logotropic DM halos decreases at large distances as
$r^{-1}$ so
that their mass is infinite. This may not be a too serious problem because
isothermal DM halos also face the same infinite mass problem while they have
often been used
to model DM halos. Actually, the logotropic or the isothermal equation of state
is only valid in the core of DM halos. In practice, this core region is
surrounded by an extended
envelope resulting from a process of (incomplete) violent relaxation
\cite{lb,wignerPH,logosf} where the density
decreases more rapidly than $r^{-1}$ and ensures a finite mass. Logotropic DM
halos have
a constant surface density (with the correct value) determined by the
fundamental constant $A$ while isothermal DM halos have not a constant surface
density unless the temperature $T$ changes from halo to halo in a rather {\it ad
hoc} manner (in the absence of a more solid justification) according to the law
$k_B T/m\propto G\Sigma_0 r_h$ \cite{modeldmB,modeldmF}.

(ii) The speed of sound $c_s$ in  logotropic DM halos increases as the
density decreases \cite{epjp,lettre,jcap,pdu,action,logosf}. This can inhibit
the formation of large scale structures by Jeans instability and
produce damped oscillations in the matter power spectrum (similar to baryonic or
acoustic oscillations) that are not observed. The Chaplygin gas faces similar
difficulties. This is actually a problem of all UDM models (see the discussion
in Sec. XVI of \cite{logosf}). These difficulites have been evidenced at
the level of linear perturbation theory \cite{sandvik}. It is not obvious if
they persist in the fully nonlinear problem \cite{logosf,zant}. This will be an
important point to
clarify in the future but it requires sophisticated numerical simulations.

We would like to conclude this paper by stressing again one of our most
striking results. We know from observations that the present universe contains
approximately $70\%$ DE and $25\%$ DM. Their ratio is $\sim 2.8$. Our line of
investigation based on the logotropic model suggests that this ratio may be
equal to
$e=2.7182818...$. This ``prediction'' is interesting in itself, independently
from the
logotropic model. It would
be of considerable interest to test this prediction with more accurate
measurements and try to understand its meaning and the strong cosmic
coincidence that it implies.

\end{document}